\documentclass[usenatbib]{mn2e}

\usepackage[toc,page]{appendix}
\usepackage{amsmath}
\usepackage[dvips]{graphicx}
\usepackage{epsfig}
\usepackage{lscape}
\usepackage{natbib}
\usepackage{amssymb}
\usepackage{amsmath}

\newcommand{\kms}{{\rm \,km\,s^{-1}}}
\newcommand{\Mpc}{\,{\rm Mpc}}
\newcommand{\kpc}{\,{\rm kpc}}

\newcommand{\Rcusp}{R_{\rm cusp}}
\newcommand{\Rfold}{R_{\rm fold}}

\newcommand{\Rein}{\theta_{\rm Ein}}

\title[Explaining lensing flux-ratio anomalies with CDM substructures]
{How well can cold-dark-matter substructures account for the observed
  lensing flux-ratio anomalies?}

\author[Xu et al.] {Dandan Xu$^{1}$\thanks{E-mail:
    xudd@astro.uni-bonn.de}, Dominique Sluse$^{1}$, Liang Gao$^{2,3}$,
  Jie Wang$^{3}$, Carlos Frenk$^{3}$, \and Shude Mao$^{2}$, Peter
  Schneider$^{1}$ \\$^{1}$Argelander-Institut f$\ddot{u}$r
  Astronomie, Universit$\ddot{a}$t Bonn, Auf dem H$\ddot{u}$gel 71,
  53121 Bonn, Germany \\$^{2}$National Astronomical Observatories,
  Chinese Academy of Sciences, Beijing, 100012, China
  \\$^{3}$Institute for Computational Cosmology, Dept. of Physics,
  University of Durham, United Kingdom }

\date{Accepted ...... Received ...... ; in original form......   }

\begin{document}
\pagerange{\pageref{firstpage}--\pageref{lastpage}} \pubyear{2012}
\maketitle
\label{firstpage}
\begin{abstract}
  Lensing flux-ratio anomalies are most likely caused by gravitational
  lensing by small-scale dark matter structures. These anomalies offer
  the prospect of testing a fundamental prediction of the cold dark
  matter (CDM) cosmological model: the existence of numerous
  substructures that are too small to host visible galaxies. In two
  previous studies we found that the number of subhalos in the six
  high-resolution simulations of CDM galactic halos of the Aquarius
  project is not sufficient to account for the observed frequency of
  flux ratio anomalies seen in selected quasars from the CLASS
  survey. These studies were limited by the small number of halos
  used, their narrow range of masses ($\sim (1-2) \times 10^{12}
  M_{\sun})$ and the small range of lens ellipticities considered. We
  address these shortcomings by investigating the lensing properties
  of a large sample of halos with a wide range of masses in two sets
  of high resolution simulations of cosmological volumes and comparing
  them to a currently best available sample of radio quasars.
  We find that, as expected, substructures do not change the flux-ratio
  probability distribution of image pairs and triples with large
  separations, but they have a significant effect on the distribution
  at small separations.  For such systems, CDM substructures can account
  for a substantial fraction of the observed flux-ratio anomalies.
  For large close-pair separation systems, the discrepancies existing
  between the observed flux ratios and predictions from smooth halo
  models are attributed to simplifications inherent in these models which
  do not take account of fine details in the lens mass distributions.

\end{abstract}

\begin{keywords}
  gravitational lensing: strong - galaxies: haloes - galaxies:
  structure - cosmology: theory - dark matter.
\end{keywords}

\section{Introduction}
In the cold dark matter (CDM) model of structure formation a large
population of dark matter subhalos is predicted to survive inside
larger ``host'' halos. In galaxies like the Milky Way, their number
vastly exceeds the number of observed satellites (about two dozen
have been discovered in the Milky Way to date). This discrepancy can
be readily understood on the basis of standard ideas on galaxy
formation \citep{Bullock2000ReionizationFixSatelliteCount,
Benson2002PhotoionizationFixSatelliteCount, Cooper2010GalaxyCDM,
Font2011GalaxySatelliteCDM, Guo2011Dwarf2cDgalaxyCDM}. The model of
\citet{Benson2002PhotoionizationFixSatelliteCount} predicted a
population of ultrafaint satellites which was subsequently
discovered in the SDSS \citep{Tollerud2008,Koposov2008MWSatelliteLF,
Koposov2009MWSatellites}. But even in this case, a large number of
subhalos too small to make galaxies, should exist. These could, in
principle, be detected from their strong gravitational lensing
effects.


The present of subhalos could be revealed by the anomalous flux
ratios seen in some multiply-lensed quasar systems. In these cases
standard parametric lens models (e.g., a singular isothermal
ellipsoid plus external shear, hereafter ``SIE+$\gamma$'') can fit
the image positions well, but not their flux ratios. This is known
as the ``anomalous flux ratio'' problem (\citealt{Kochanek91}).

A number of possible solutions to the problem have been considered.
For example, some of the flux-ratio anomalies could be accommodated
by adding higher order multipoles to the ellipsoidal potential of
the lensing galaxy.  However, as the required amplitudes would be
unreasonably larger than typically observed in galaxies and halo
models, such a solution does not seem plausible
(\citealt{EW2003,KD2004,CongdonKeeton2005,Yoo2006}).

Propagation effects in the interstellar medium, such as galactic
scintillation and scatter broadening, could also cause anomalous
flux ratios. If so, one would expect a strong wavelength dependence
of the anomalies measured at radio wavelengths, which has not been
seen (\citealt{Koopmans2003JVASCLASS,KD2004}). Moreover, propagation
effects cannot explain the fact that observed saddle
(negative-parity) images often appear to be fainter than predicted
by the standard lens models (e.g., \citealt{SW2002apj,KD2004}).

The most promising explanation is that the anomaly is due to
perturbations from small-scale structures hosted by lensing
galaxies. Microlensing, which refers to lensing by stars, has been
identified as a source of the flux-ratio anomalies observed at
optical wavelengths for the multiple images of background quasars
(\citealt{Wozniak2000Q2237ml, Sluse2006J1131, Keeton2006sdss0924ml,
  Morgan2004WFIi2026ii2033, Morgan2006sdss0924ml, Morgan2008PG1115,
  Vuissoz2008WFI2033}). In this case, the Einstein radius of a star in
a foreground lensing galaxy is comparable to the size of the optical
emission from the accretion disk of a background quasar; the
magnification effect (on scales of the Einstein radius) alters the
flux ratio between the images.

However, this is not the case for radio emission, which comes from a
region much larger than the accretion disk. Any magnification due to
foreground stars would be smeared out within the whole radio image
zone and thus become insignificant. Therefore microlensing is not a
likely explanation for the flux ratio anomalies measured at radio
wavelengths (e.g., \citealt{KD2004}).

\citet{MS1998mn} first realized that dark matter substructures (on
scales smaller than image separations $\sim 1\arcsec$ for typical
lens and source redshifts) could explain the radio flux-ratio
anomaly in B1422+231. Later studies showed that the presence of
substructures in lensing galaxies can also explain the observed
tendency for the brightness of saddle image to be suppresed
(\citealt{SW2002apj, Keeton2003SaddleImages,KD2004}). Lensing
subhalos has therefore emerged as one of the most convincing
explanations for the radio flux-ratio anomalies. Such an explanation
has important implications for cosmology since it provides a
straightforward and very direct test of the CDM model.


There are about a dozen studies that use $N$-body simulations to
test if the predicted CDM substructures are sufficiently abundant to
explain the frequency of anomalous lenses in currently available
samples. However, no consensus has emerged. While some of the
studies suggest consistency between the CDM theory and observations
(e.g., \citealt{DK2002, BS2004aa,
DoblerKeeton2005,Metcalf2010FluxAnomaly}), others, including those
by us (\citealt{Dandan09AquI,Dandan2010AqII}), find otherwise (e.g.,
\citealt{MaoJing04apj, AB06mn, Maccio2006b, Maccio2006,
  Chen2011CuspViolation}).

\citet{Metcalf2010FluxAnomaly} pointed out that the different
conclusions between their study, which finds consistency between
theory and data, and others, like ours, which do not
(\citealt{Dandan09AquI,Dandan2010AqII}) could be due to the
following: (1) as the number density of substructures is small near
the radii where multiple images form, relying on only a few
projections of a small number of high-resolution halos, as we did,
could produce biased results due to halo-to-halo variations; (2) as
flux ratios are quite sensitive to the ellipticity of the main lens
(\citealt{KGP2003apj}), our restriction to a relatively small
ellipticity (axis ratio = 0.8), instead of the full range of
ellipticities in the main lens models, could also skew the results.

There are other factors that could weaken the conclusion in our
studies. For example, the cosmological simulations that we have used
contain only dark matter but no baryons, and this might change the
subhalo survival rate and thus their spatial distributions. This
effect is negligible for substructures in galaxies which have very
large (or infinite) mass-to-light ratios but could be important in
groups and clusters. Unfortunately, current simulations are not yet
sufficiently large or accurate to investigate this problem. We will
have to rely on using the next-generation hydrodynamical
cosmological simulations to accommodate such effects.

Another limitation of our previous work is that we focused
exclusively on the six Milky Way-sized dark matter halos from the
Aquarius project (\citealt{volker08Aq}). However, massive
ellipticals galaxies, which comprise 80\%-90\% of observed lenses
\citep{KKF1998,KFI2000,RKF2003} are more likely to occur in
group-sized halos which are ten times more massive.
Since the subhalo abundance increases rapidly with increasing host
halo mass \citep[e.g.][]{DeLucia2004CDMSub, Gao2004b,
Zentner2005SubhaloPopulation, Wang2012MassiveHostMoreSub}, the
subhalo population in the Aquarius simulations may not fairly
represent that in group-sized dark matter halos and this could have
led us to underestimate the probability of flux-ratio anomalies.


In this paper we address most of the issues mentioned above. We use
two sets of high-resolution cosmological CDM simulations - from the
Aquarius \citep{volker08Aq} and Phoenix \citep{Gao12Phoenix} - to
attempt to answer the question of how well do cold-dark-matter
substructures account for the observed lensing flux-ratio anomalies.
The paper is organized as follows: In Sect.2, we review the generic
relations in cusp (Sect.2.1) and fold (Sect.2.2) lenses. In
Sect.2.3, we present our observational sample of eight lenses, all
of which have radio measurements for both cusp and fold relations.
In Sect.3, we present a method to represent numerically massive
elliptical lenses and their subhalo populations.  To model the
potentials of a general host lens population, in Sect.3.1, we use a
technique similar to that of Keeton et al. (2003). In Sect.3.2, we
rescale both the Aquarius and Phoenix simulations to group-sized
dark matter halos, add the rescaled subhalos to the smooth lens
potentials, and estimate their perturbation effects on the flux
ratios of quadruply-lensed quasar images.
The resulting flux-ratio probability distributions are presented in
Sect.4, where comparisons are made to observations and to previous
studies. Due to the limitations of generalized smooth lens
potentials and image configurations, we have further carried out a
detailed investigation of each system in the currently available
sample; results presented in Sect.5. A discussion and our
conclusions are given in Sect.6.

The cosmology we adopt here is the same as that for both sets of
simulations that we have used in this work, with a matter density
$\Omega_{\rm m}$ = 0.25, cosmological constant $\Omega_{\Lambda}$ =
0.75, Hubble constant $h=H_0/(100\kms\Mpc^{-1})=0.73$ and linear
fluctuation amplitude $\sigma_8=0.9$. These values are consistent
with cosmological constraints from the WMAP 1- and 5-year data
analyses (\citealt{WMAP-1, WMAP-5}), but differ from Planck 2013
results (\citealt{Planck2013}), where $h=0.67$ and
$\sigma_8=0.829\pm0.012$. However, we do not expect these
differences in cosmological parameters to have any significant
consequences for our conclusions.

\section{Generic relations in cusp lenses and fold lenses}

\begin{figure}
\centering
\includegraphics[width=7cm]{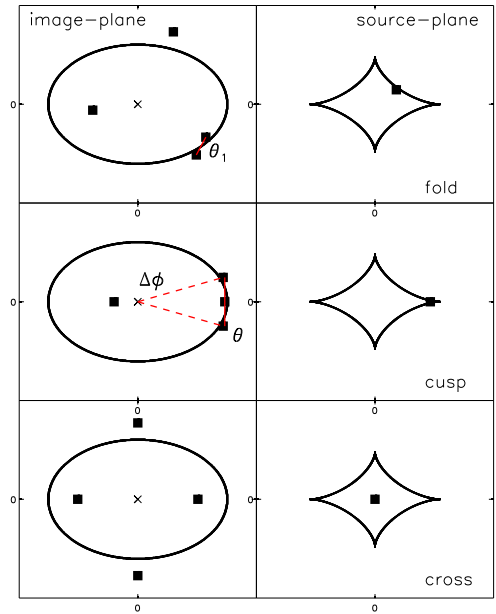}
\caption{Three basic image configurations: fold (top), cusp
(middle), and cross (bottom), with respect to the tangential
critical curves in the image plane (on the left), and corresponding
source positions with respect to the central caustics in the source
plane (on the right). The image separation $\theta_1$ of a close
pair is labeled for the fold configuration; image opening angle
$\Delta\phi$ and separation $\theta$ of a close triplet are labeled
for the cusp configuration.} \label{fig:CuspFoldCrossConfiguration}
\end{figure}

There are three generic configurations of four-image lenses (see
Fig.\ref{fig:CuspFoldCrossConfiguration}): (1) a source located near
a cusp of the tangential caustic will produce a ``cusp''
configuration, where three images form close to each other around
the critical curve on one side of the lens; (2) a source located
near the caustic and between two adjacent cusps will produce a
``fold'' configuration, where a pair of images form close to each
other near the critical curve; (3) a source located far away from
the caustic, i.e., in the central region of the caustic, will
produce a ``cross'' configuration, where all four images form far
away from each other and away from the critical curve. Close triple
images in cusp lenses and close pair images in fold lenses are the
brightest images among the four, as they form close to the
(tangential) critical curve.

There are some universal magnification relations for the triple and
pair images in cusp and fold systems under smooth lens potentials.
These generic relations assist one to identify small-scale
perturbations via identifying violations to these generic
magnification relations, without requiring detailed lens modeling
for individual systems.

\subsection{The cusp relation and the violation due to small-scale
  structures}

In any smooth lens potential that produces multiple images (of a
single source) of a cusp configuration, a specific magnification ratio
(i.e., also the flux ratio) of the image triplet will approach zero
asymptotically, as the source approaches a cusp of the tangential
caustic. This is known as the ``cusp relation'' (\citealt{BN1986apj,
  SW1992aa, Zakharov1995AA, KGP2003apj}), mathematically defined
as:
\begin{equation}
  \Rcusp \equiv \frac{|\mu_A + \mu_B +
    \mu_C|}{|\mu_A|+|\mu_B|+|\mu_C|} \rightarrow 0 ~~~
  (\Delta\beta \rightarrow 0),
\label{eq:Rcusp}
\end{equation}
where $\Delta\beta$ is the offset between the source and the nearest
cusp of the caustic, $\mu_{A, B, C}$ denote the triplet's
magnifications, whose signs indicate image parities.

Because $\Delta\beta$ cannot be directly measured, we therefore
follow the practice of Keeton et al. (2003), using $\Delta\phi$ and
$\theta/\Rein$ to quantify a cusp image configuration. As labeled in
Fig.\ref{fig:CuspFoldCrossConfiguration}, $\Delta\phi$ is defined as
the angle between the outer two images of a triplet, measured from
the position of the lens centre; $\theta/\Rein$ is the maximum image
separation among the triplet, normalized by the Einstein radius
$\Rein$. In general, when the source moves towards the nearest cusp,
both $\Delta\phi$ and $\theta/\Rein$ will decrease to zero.


Small-scale structures, either within the lens or projected by chance along
the line of sight, will perturb the lens potential and alter fluxes of one or
more images. 
In this case, $\Rcusp$ will become unexpectedly large. The cusp
relation, i.e., $\Rcusp \rightarrow 0$ when $\Delta\beta\rightarrow
0$, will then be violated (e.g., see Fig.2 of Xu et al. 2012 for an
illustration).

\subsection{The fold relation and the violation due to small-scale
  structures}

For an image pair in a fold configuration produced by any smooth lens
potential, there is also a generic magnification relation, namely the
``fold relation'' (\citealt{BN1986apj, SW1992aa, SchneiderBook1992,
  PettersBook2001}). In this paper, we take the form as in Keeton et
al. (2005):
\begin{equation}
  \Rfold \equiv \frac{|\mu_{\rm min}|-|\mu_{\rm sad}|}{|\mu_{\rm min}|+|\mu_{\rm sad}|}
  \rightarrow 0 ~~~ (\Delta\beta \rightarrow 0),
\label{eq:Rfold}
\end{equation}
where $\Delta\beta$ is the offset of the source from the fold
caustic, $\mu_{\rm min, sad}$ denote magnifications of the minimum
($\mu>0$) and saddle ($\mu<0$) images. To quantify a fold image
configuration, similar to the practice of Keeton et al. (2005), we
use $\theta_1/\Rein$ to indicate how close the pair of images are.
As labelled in Fig.\ref{fig:CuspFoldCrossConfiguration},
$\theta_1/\Rein$ is defined as the separation, in unit of the
Einstein radius $\Rein$, between the saddle image and the nearest
minimum image. When small-scale structures are present, $\Rfold$
will also become unexpectedly large; the fold relation, i.e.,
$\Rfold \rightarrow 0$ when $\Delta\beta\rightarrow 0$, will then be
violated.

We can study small-scale structures by investigating the resulting
violations to the cusp and fold relations in extreme systems where
$\Delta\beta\sim0$. However, the detected number of such systems is
zero. For observed lenses, $\Delta\beta\neq 0$, and the exact values
of $\Rcusp$ and $\Rfold$ depend on $\Delta\beta$ and on the lens
potentials. Without detailed lens modelling, we can identify cases
of violations as outliers of some general distributions of $\Rcusp$
and $\Rfold$ for smooth lenses. A series of comprehensive and
detailed studies on this topic have been carried out by e.g., Keeton
et al. (2003, 2005), whose methods are largely followed in this work
(see Sect.3.1).

\subsection{A sample of cusp and fold lenses}

As adding CDM substructures to smooth lens potentials would change
the distributions of $\Rcusp$ and $\Rfold$, resulting in anomalous
flux ratios, we can compare simulation results to a reasonable
observational sample to see how well the CDM substructure models can
explain observations. There are more than twenty lenses with
$\Rcusp$ and $\Rfold$ measured at optical, near-infrared (NIR) and
radio wavelengths (CfA-Arizona Space Telescope Lens Survey,
http://cfa-www.harvard.edu/castles). Nearly half of them show
anomalous flux ratios in the sense that the measured $\Rcusp$ and
$\Rfold$ cannot be reproduced by the lens models that best fit their
image positions (Keeton et al. 2003, 2005).

As the fluxes measured at optical and NIR (\citealt{SluseMIR2013})
wavelengths can be significantly affected by microlensing and dust
extinction, we will only take those systems with $\Rcusp$ and $\Rfold$
measured at radio wavelengths.

In our previous studies, we used all (five) existing radio lenses that
have triplet image opening angles $\Delta\phi \leqslant
120^{\circ}$. All of them have surprisingly large $\Rcusp$ values that
cannot be reproduced by the best-fit lens models. In this work, we
take all radio quads (four distinct point-like images from one single
source) in either cusp or fold configurations that have both $\Rcusp$
and $\Rfold$ measurements. This forms a sample of a total of eight
lenses; three ``cusp'', and five ``fold''.

\begin{figure*}
\centering
\includegraphics[width=5.5cm]{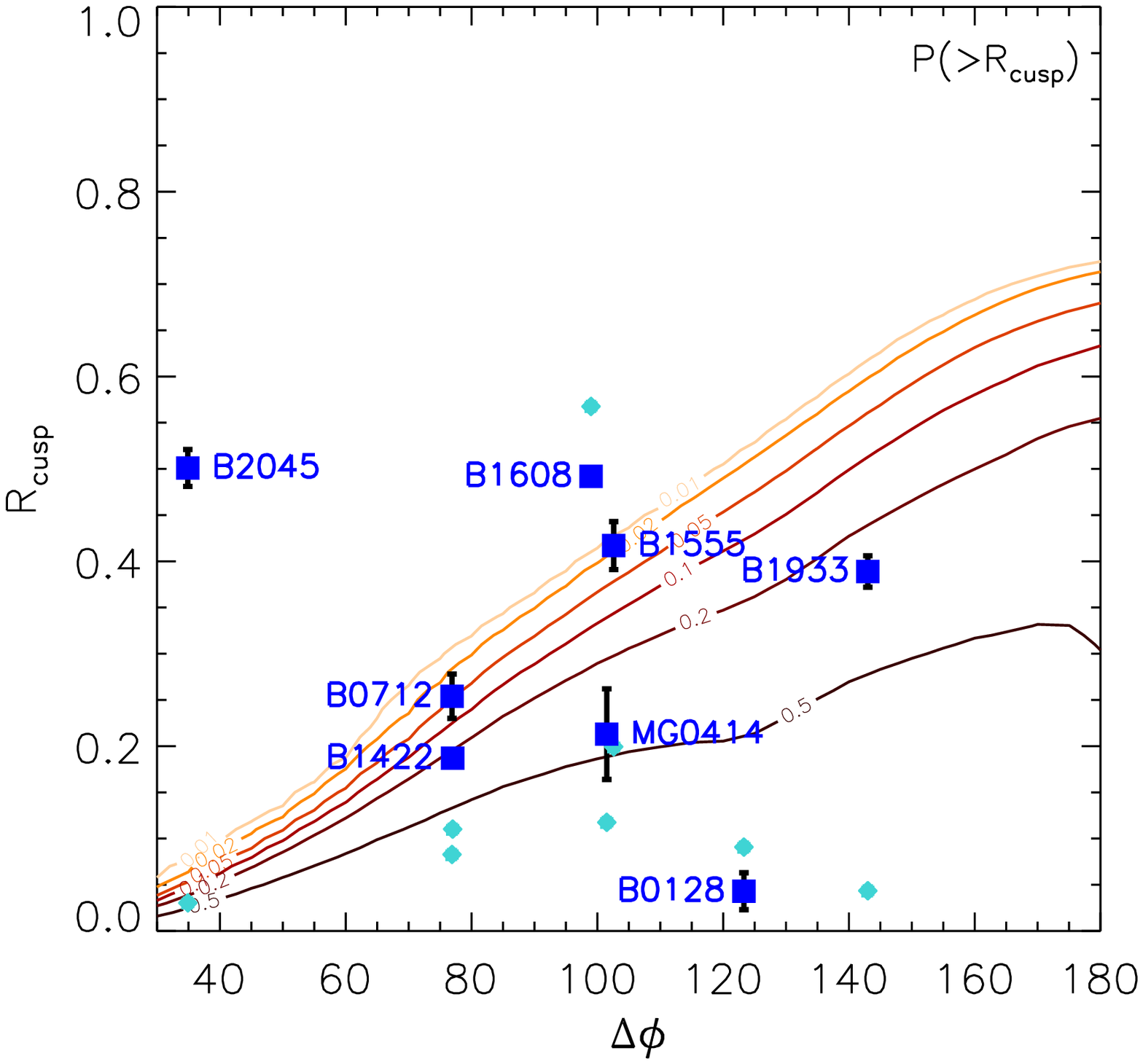}
\includegraphics[width=5.5cm]{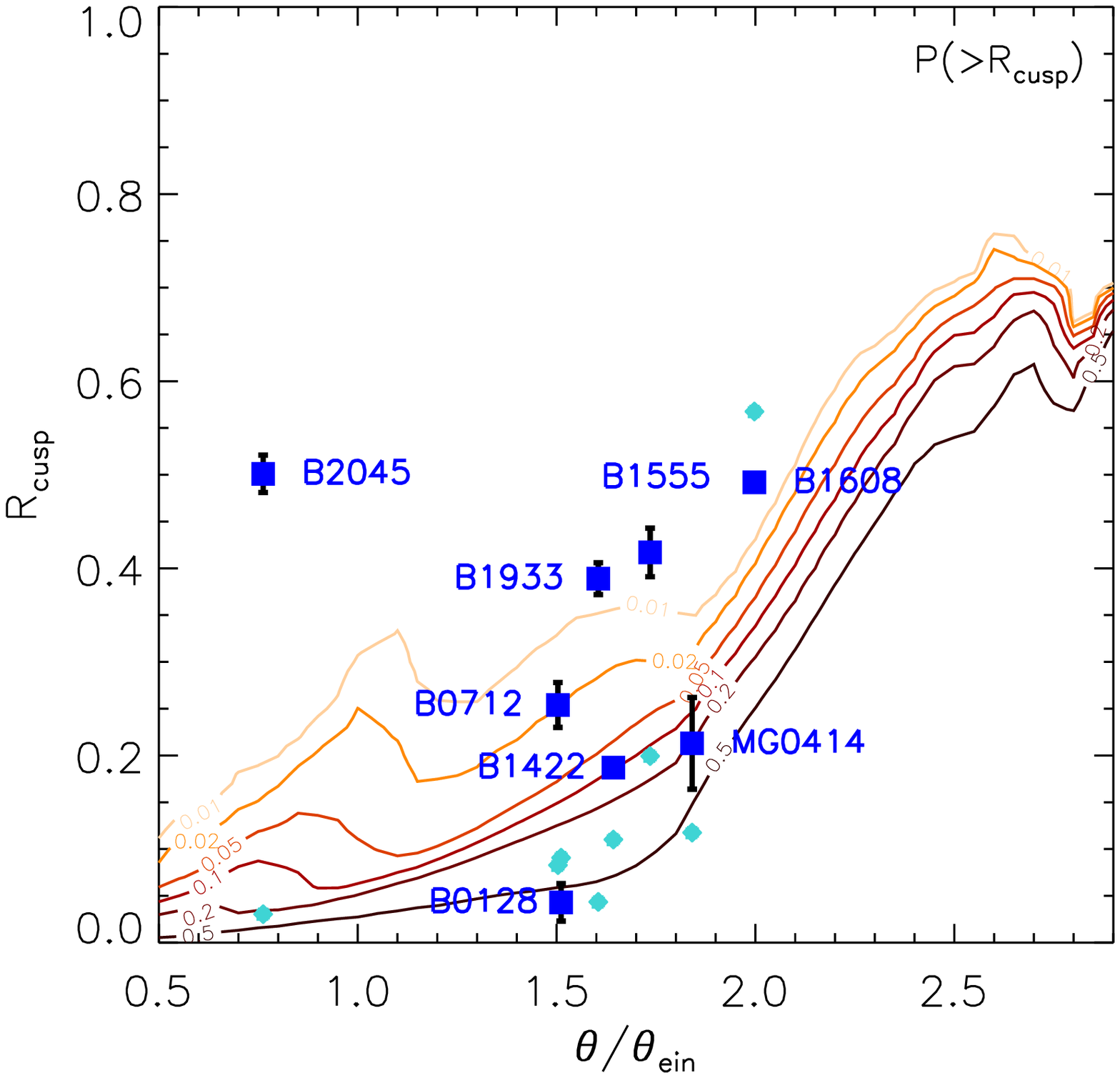}
\includegraphics[width=5.5cm]{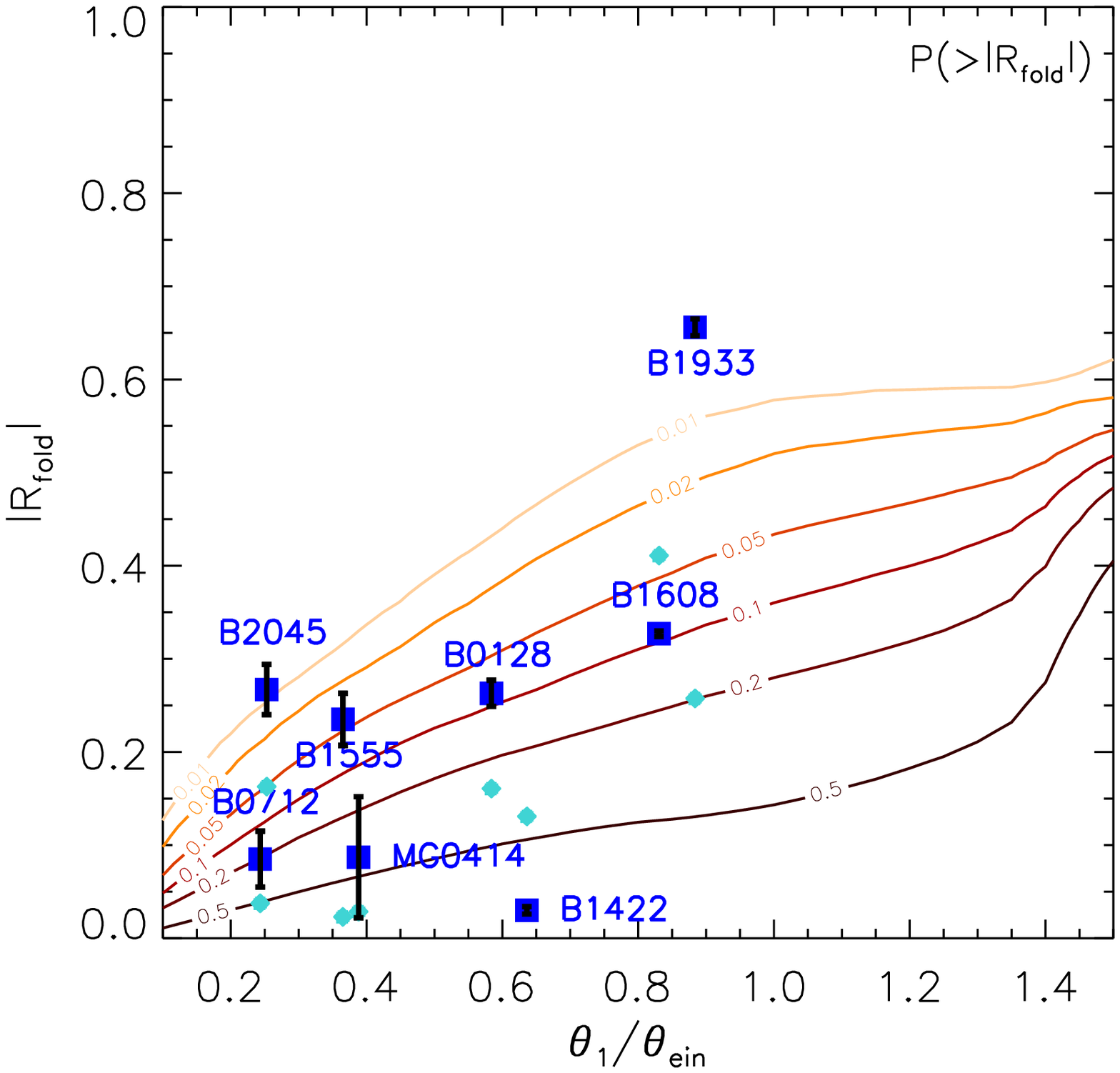}\\
\includegraphics[width=5.5cm]{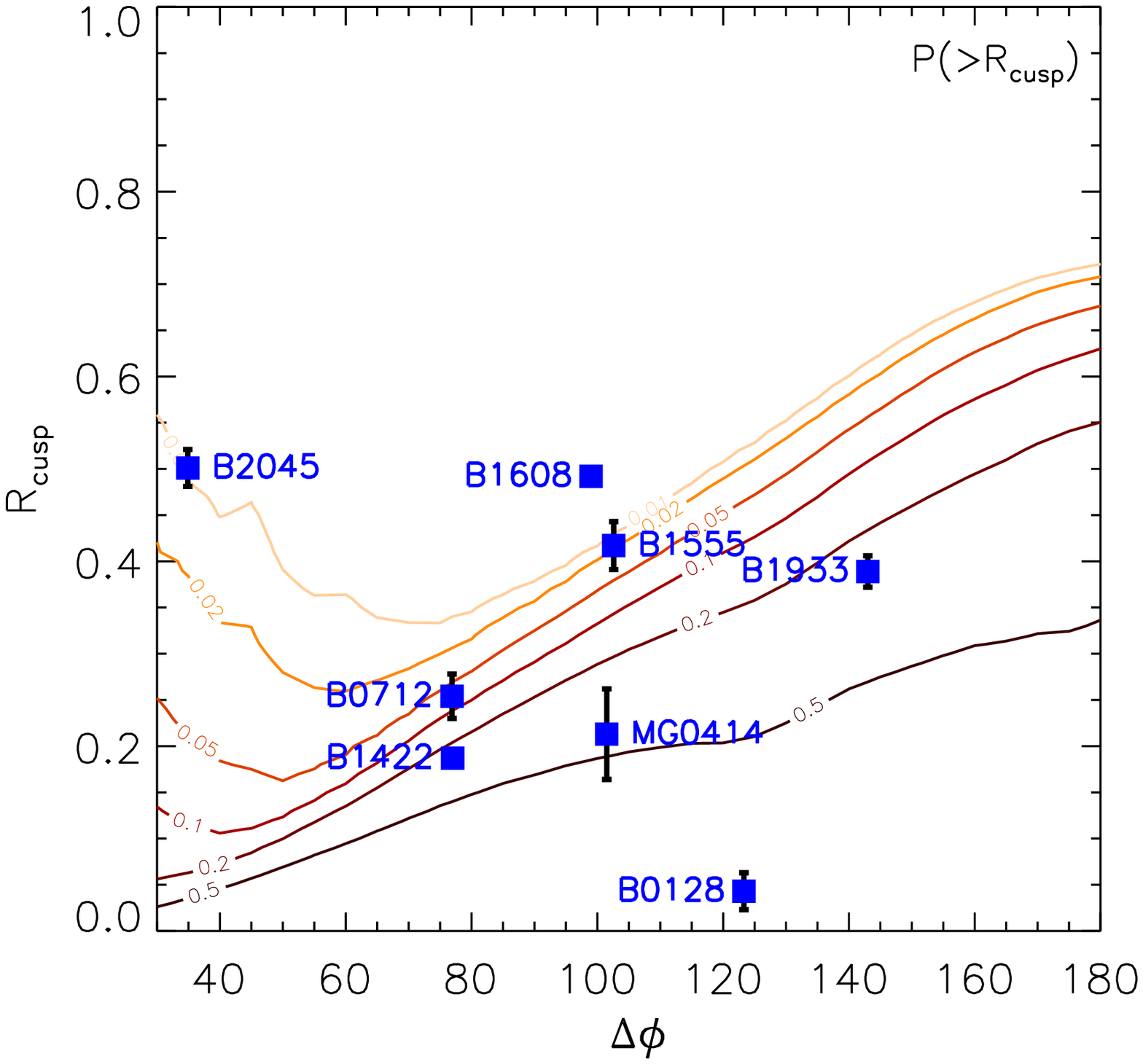}
\includegraphics[width=5.5cm]{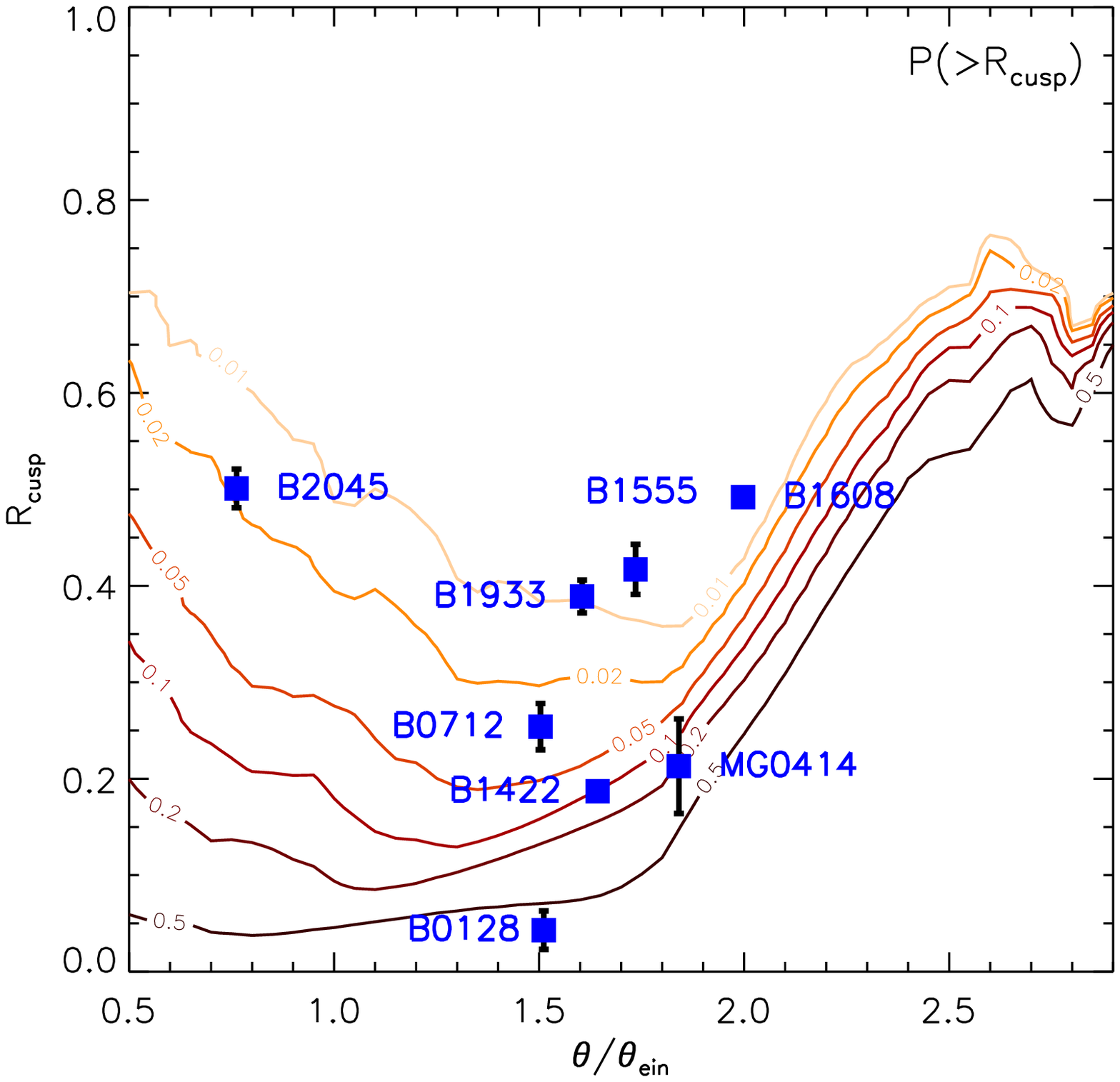}
\includegraphics[width=5.5cm]{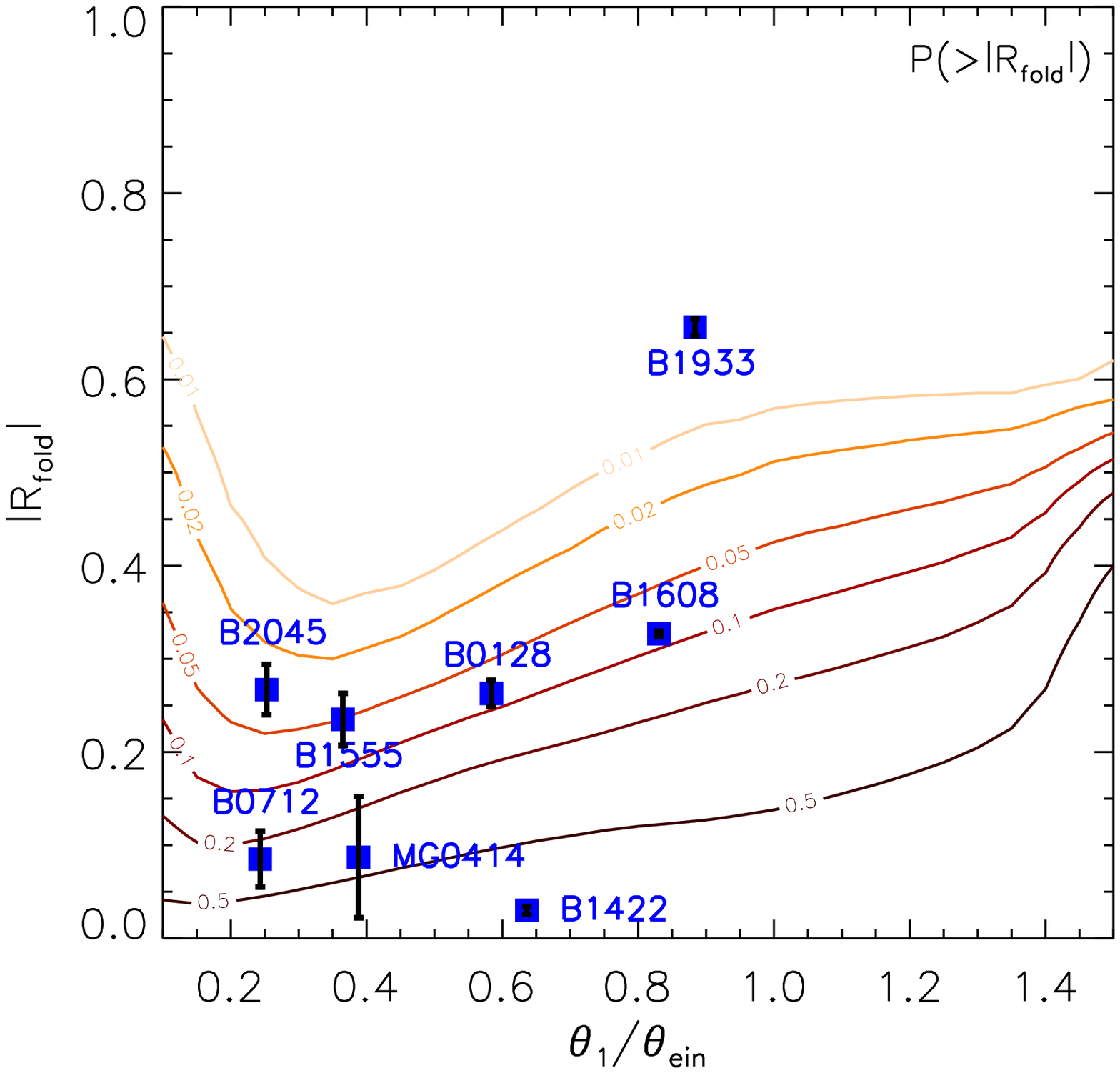}\\
\includegraphics[width=5.5cm]{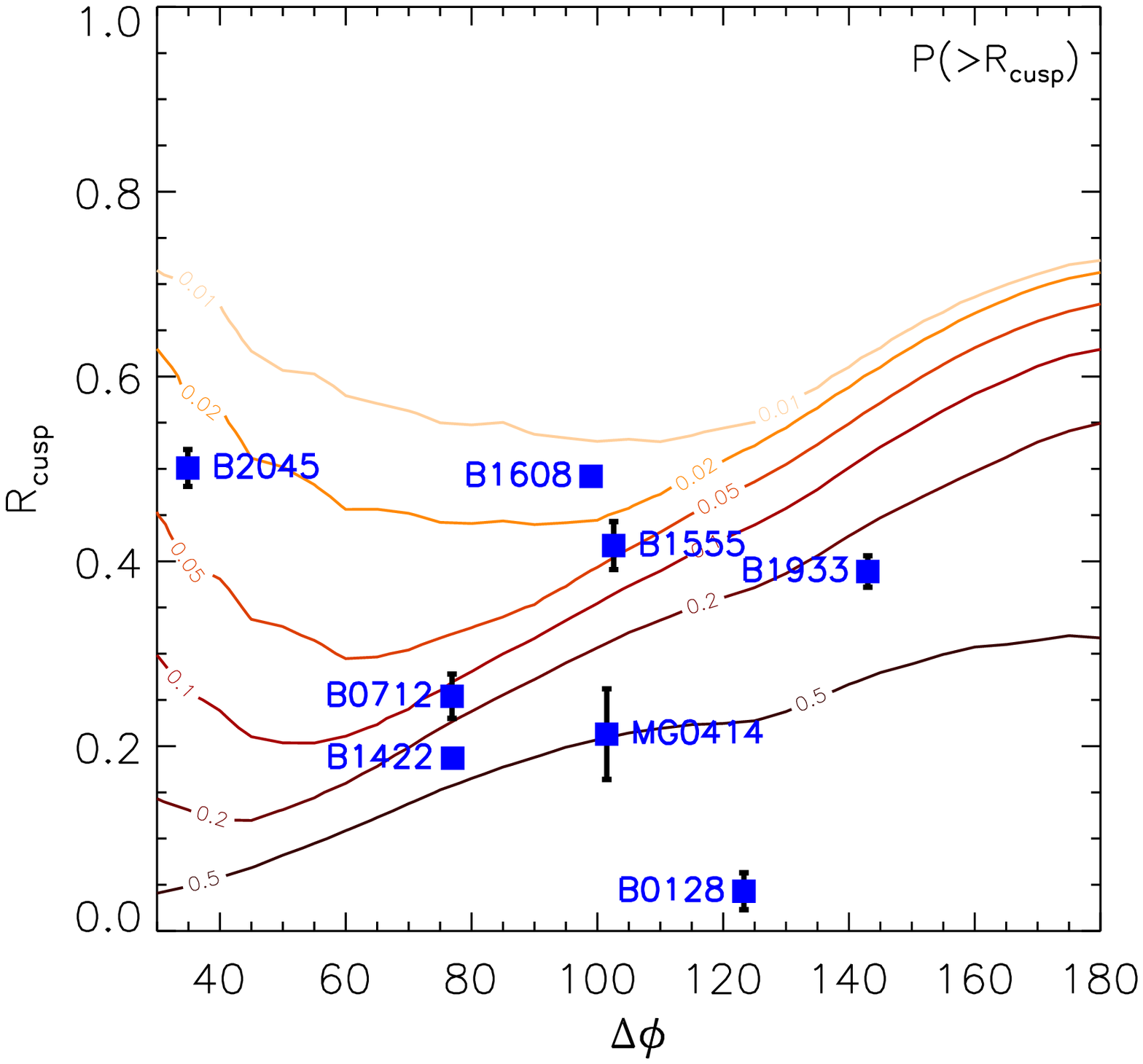}
\includegraphics[width=5.5cm]{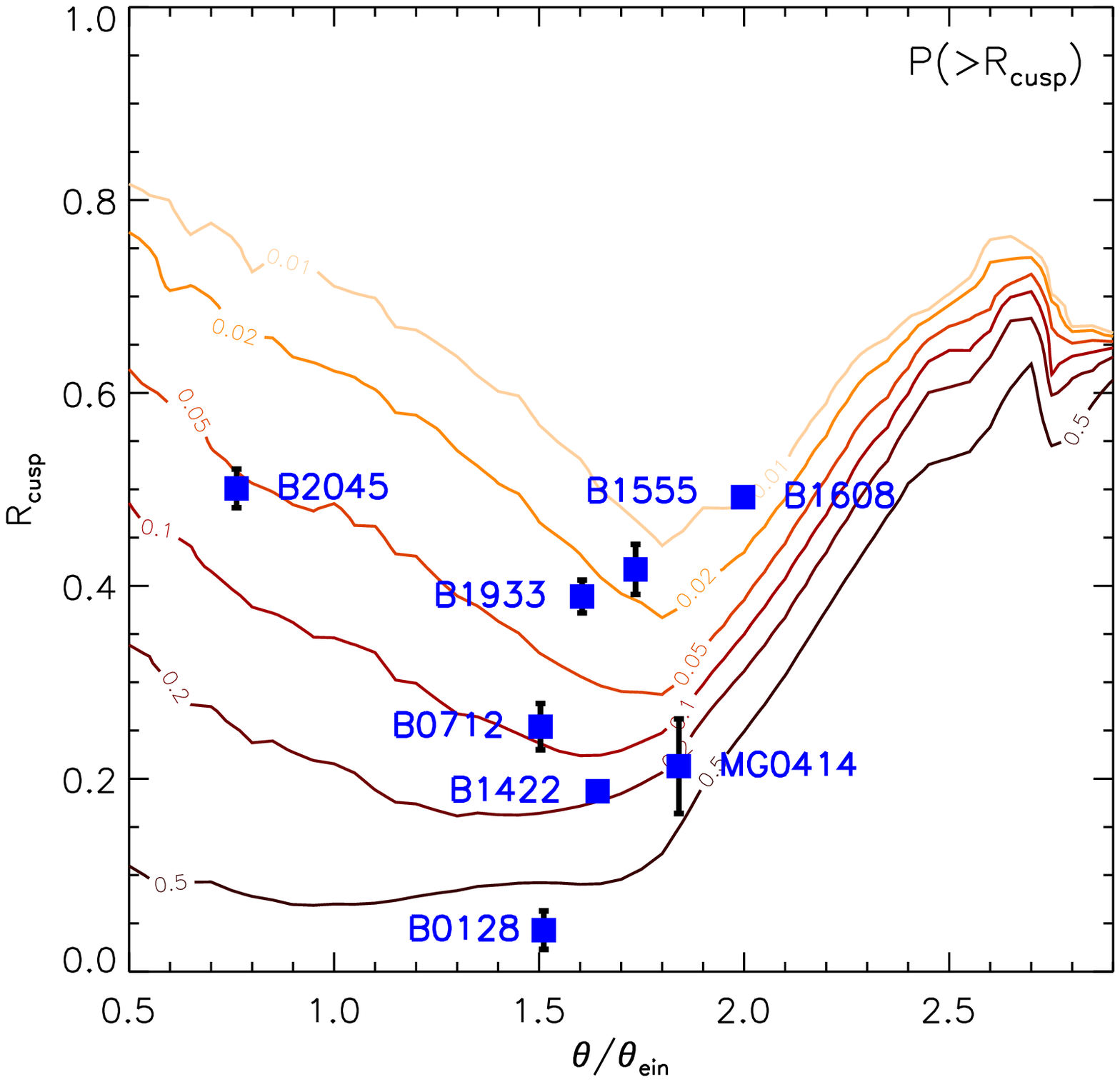}
\includegraphics[width=5.5cm]{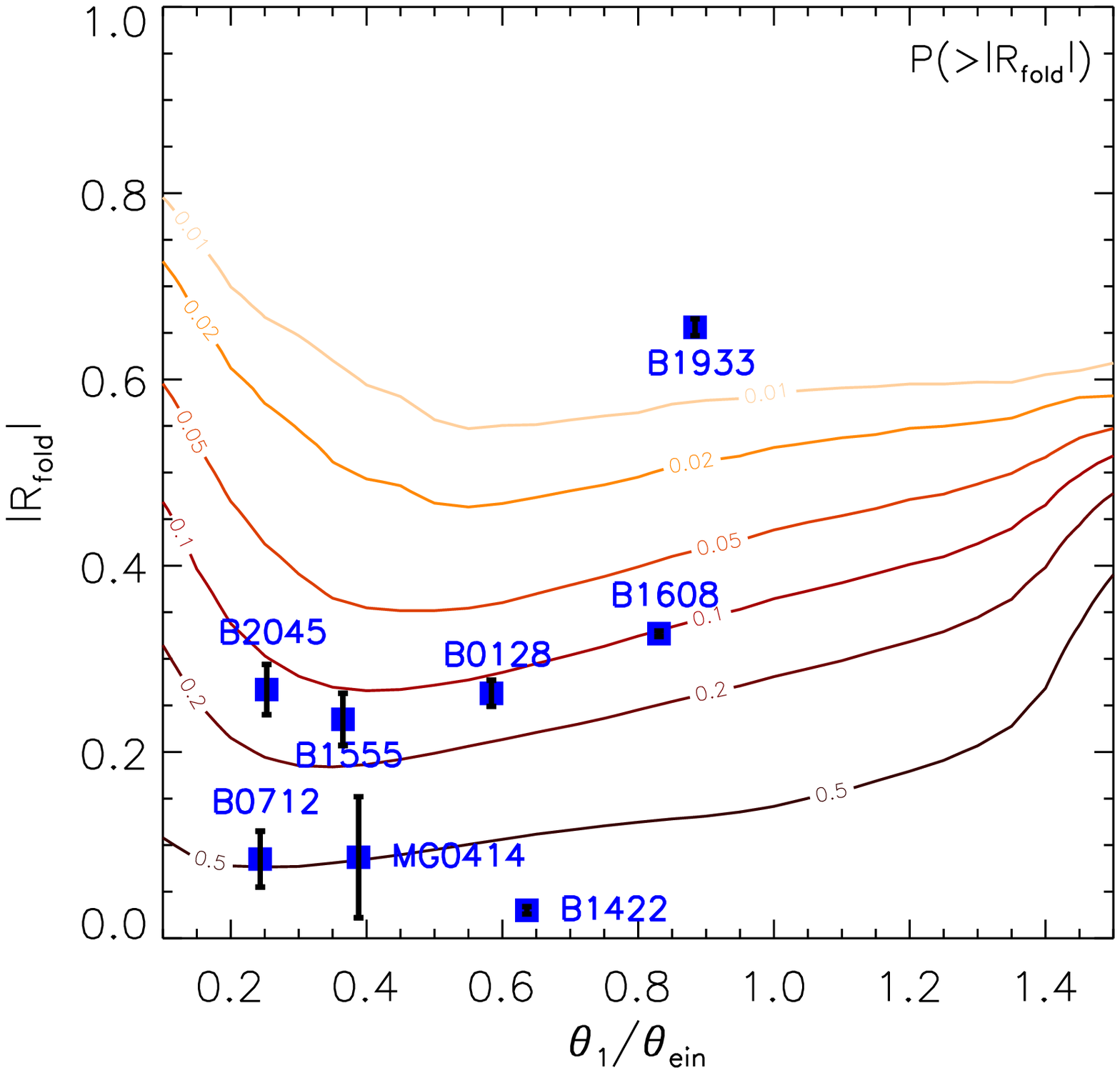}\\
\caption{Probability contour maps of conditional probabilities
  $P(>\Rcusp\mid \Delta\phi)$ (left column), $P(>\Rcusp\mid
  \theta/\Rein)$ (middle column) and $P(>|\Rfold|\mid \theta_1/\Rein)$
  (right column). Contour levels of 1, 2, 5, 10, 20 and 50 per cent
  (from light to dark) are plotted. Top: singular isothermal
  ellipsoidal potentials with axis ratio $q$ and higher-order
  perturbation amplitudes $a_m$ drawn from 847 observed
  galaxies (Hao et al. 2006), plus randomly oriented external shear.
  Middle: smooth potentials (as above) plus perturbations from a
  simulated subhalo population hosted by a Milky Way-sized halo of
  $M_{200}=10^{12}h^{-1}M_{\odot}$. Bottom: smooth potentials (as for
  the top panel) plus perturbations from a simulated subhalo
  population hosted by a group-sized halo of
  $M_{200}=5\times10^{13}h^{-1}M_{\odot}$. More than $5\times10^6$
  realizations have been calculated for each case.
  Measured and model predicted flux ratios ($\Rcusp$ and $|\Rfold|$) of eight
  observed lenses are plotted as blue squares and cyan diamonds,
  respectively; measurement errors are also given.
  These distributions do not vary with the way that
  data are binned when using a reasonable
  range of bin sizes.} \label{fig:CuspViolationTotal}
\end{figure*}

For each of the eight observed lenses, we take the basic image
configuration measurements, namely, $\Delta\phi$, $\theta/\Rein$ and
$\theta_1/\Rein$, as well as the measured and model predicted flux
ratios of $\Rcusp$ (for the closest triple images) and $\Rfold$ (for
the closest saddle-minimum image pairs), listed in Table
\ref{tab:ObsSample-Cusp}. It can be seen that discrepancies at
different levels exist between the observed flux ratios and the
predictions from best-fit smooth models.




\begin{table*}
\centering \caption{Observed lenses with measurements of $\Rcusp$
and $\Rfold$ for the close triple images:}
\label{tab:ObsSample-Cusp}
\begin{minipage} {\textwidth}
\begin{tabular}[b]{l|c|c|c|c|c|c|c}\hline\hline
  ~~~~Lens & ~~~Type~~~& ~~~$\Delta\phi$(${^\circ}$)~~~ & ~~~$\theta/\Rein$~~~ &
  ~~~$\Rcusp$~~~& ~~~$\theta_1/\Rein$~~~ & ~~~$\Rfold$~~~
  & ~~References~~ \\\hline
B0128+437     & fold  & 123.3 & 1.511  & 0.043$\pm$0.020 (0.090) & 0.584 & 0.263$\pm$0.014 (0.161) & 1, 2 \\
MG0414+0534   & fold  & 101.5 & 1.841  & 0.213$\pm$0.049 (0.118)  & 0.388 & 0.087$\pm$0.065 ($-$0.029) & 3, 4, 5, 6 \\
B0712+472 & cusp & 76.9  & 1.503  & 0.254$\pm$0.024 (0.083) & 0.243 & 0.085$\pm$0.030 ($-$0.037) & 1, 7, 8, 9\\
B1422+231     & cusp  & 77.0  & 1.643  & 0.187$\pm$0.004 (0.110) & 0.636 & $-$0.030$\pm$0.004 ($-$0.131) & 1, 10, 11, 3\\
B1555+375     & fold  & 102.6 & 1.735  & 0.417$\pm$0.026 (0.199) & 0.365 & 0.235$\pm$0.028 (0.023) & 1, 12 \\
B1608+656     & fold  & 99.0  & 1.997  & 0.492$\pm$0.002 (0.568) & 0.831 & 0.327$\pm$0.003 (0.411) & 13, 14 \\
B1933+503     & fold  & 143.0 & 1.605  & 0.389$\pm$0.017 (0.040)    & 0.884 & 0.656$\pm$0.009 (0.257) & 15, 16, 17 \\
B2045+265     & cusp  & 34.9  & 0.762  & 0.501$\pm$0.020 (0.030) & 0.253 & 0.267$\pm$0.027 ($-$0.163) & 1, 9, 18, 19 \\
\hline
\end{tabular}
\\ Notes: the quoted $\Rcusp$ and
$\Rfold$ are measurements at radio wavelengths; their uncertainties
are derived from the uncertainties in flux measurements. Values in
brackets are predicted by the best-fit lens model, see our Sect.5.1.
References: (1) \citealt{Koopmans2003JVASCLASS}; (2)
\citealt{Phillips2000B0128}; (3) \citealt{FIK1999}; (4)
\citealt{Lawrence1995MG0414}; (5) \citealt{Katz1997MG0414}; (6)
\citealt{Ros2000MG0414}; (7) \citealt{Jackson98B0712}; (8)
\citealt{Jackson2000}; (9) Cfa-Arizona Space Telescope Lens Survey
(CASTLES, see http://cfa-www.harvard.edu/castles); (10)
\citealt{Impey1996B1422}; (11) \citealt{Patnaik1999B1422}; (12)
\citealt{Marlow1999B1555}; (13)
\citealt{KoopmansFassnacht1999B1608}; (14)
\citealt{Fassnacht1996B1608}; (15) \citealt{Cohn2001B1933}; (16)
\citealt{Sykes1998B1933}; (17) \citealt{Biggs2000B1933}; (18)
\citealt{Fassnacht1999B2045}; (19) McKean et al. 2007; .
\end{minipage}
\end{table*}

In order to see how observations would compare to smooth model
predictions and how frequently they are expected to violate smooth
predictions, we produce expected flux-ratio probability
distributions with a total of $5\times10^6$ realizations for a
generalized smooth lens population, which are modelled by isothermal
ellipsoidal potentials with axis ratios and high-order multipole
perturbations drawn from 847 observed galaxies (Hao et al. 2006)
plus external shear (see Sect.3.1 for more details).

The results are presented in the top panels of
Fig.\ref{fig:CuspViolationTotal}, where probability contours of
$P(>\Rcusp\mid\Delta\phi)$, $P(>\Rcusp\mid\theta/\Rein)$ and
$P(>|\Rfold|\mid\theta_1/\Rein)$ are plotted. A small (large)
probability $P$ means that it is less (more) likely for a flux ratio,
either $\Rcusp$ or $|\Rfold|$, to be larger than a given value, at a
given image configuration, described by $\Delta\phi$, $\theta/\Rein$
or $\theta_1/\Rein$.

On top of the probability contours in
Fig.\ref{fig:CuspViolationTotal}, measured $\Rcusp$ and $|\Rfold|$ of
the eight lenses in our sample are plotted as blue squares, together
with measurement errors.
Flux ratios predicted by the lens model that best fits image
positions are also given, plotted as cyan diamonds. As can be seen,
the majority of model predicted flux ratios satisfy $P\gtrsim20\%$,
while most of the measured flux ratios have probabilities well below
such a percentage, suggesting such a smooth lens population on its
own is unlikely to account for most of the observed flux ratio
anomalies.


\section{Realization of a massive elliptical lens and its subhalo population}

In order to see how CDM substructures could affect the flux-ratio
probablity distributions and how well they can account for the
observed flux-ratio anomalies in our sample, we add simulated CDM
subhalo populations to generalized smooth lens potentials and
predict the flux-ratio probablity distributions in the presence of
substructures. In this section, we present the method of the
numerical realization of massive elliptical lenses (Sect.3.1) and
their substructure populations (Sect.3.2).

\subsection{Smooth lens model}

To model the main lens halo (which is responsible for producing
multiple images), we adopt Keeton et al. (2003) smooth lens
modelling, which allows us to predict generic distributions for the
cusp and fold relations and to see how well the observational sample
follows such distributions in the absence of CDM substructures.


\citet{KGP2003apj,KGP2005Fold} studied the generic flux-ratio
relations in cusp and fold lenses and showed that they have a weak
dependence on the radial profile (from point mass to isothermal) of
the lens potential, but are sensitive to ellipticity $e$($\equiv1-q$,
where $q$ is the axis ratio), higher-order multipole amplitude $a_m$
and external shear $\gamma_{\rm ext}$. In this work, we use a
generalized isothermal ellipsoidal profile with an Einstein radius of
1.0$\arcsec$ and also take into account the three aspects
above. Detailed lens modelling is given in the Appendix.



For choosing $e$ and $a_m$, we use results from \citet{HaoMao2006},
who measured ellipticities and higher-order multipoles ($m=3,~4$) of
847 Sloan galaxies. The mean and scatter of these shape parameters
($\bar{e}=0.23$, $\sigma_e=0.13$, $\bar{a}_3=0.005$,
$\sigma_{a_3}=0.004$, $\bar{a}_4=0.010$, $\sigma_{a_4}=0.012$) are
comparable to the values reported for the galaxy samples used in
Keeton et al. (2003, 2005).

We note that by using the observed galaxy morphology distributions,
we implicitly assume that the shape of dark matter distribution
follows baryons in the inner parts of the halo where strong lensing
occurs. This has been supported by lensing observations from e.g.,
\citet{Koopmans2006apj} and \citet{Sluse2012COSMOGRAIL25}.

It is also worth noting that although we draw shape parameters ($e$
and $a_m$) for lens modelling from a galaxy sample at lower
redshifts ($z<0.2$), as addressed in Keeton et al. (2003, 2005),
such distributions are not expected to be significantly different
from those of the observed lensing galaxies at intermediate
redshifts; observations have shown no significant evolution in the
mass assembly history of early-type galaxies since $z\approx1$
(\citealt{Thomas2005,Koopmans2006apj}).



Standard lens modelling would also include external shear, which is
required to account for the lensing effect from the lens
environments (e.g., \citealt{KKS1997externalshear}, Sluse et al.
2012). We follow the practice of Keeton et al. (2003), assuming a
lognormal distribution of $\gamma_{\rm ext}$ with a median of 0.05
and $\sigma_{\gamma}=0.2$ dex.

To add simulated CDM subhalos to the generalized host lens
potentials, we take 3600 different projections of subhalo
distributions, and add each of them to one realization of the host
lens potential. For any given host lens potential, to maintain the
possible correlation between ellipticities and high-order
multipoles, we draw the combination of measured ($e$, $a_3$, $a_4$)
from the observed galaxy sample of Hao et al. (2006). We then add a
randomly orientated external shear. For each constructed lens
potential, we carry out standard lensing calculations, finding
images for source positions that are uniformly distributed inside
the tangential caustic with a number density of 20000 per arcsec$^2$
in the source plane. This naturally ensures that each realization
will be weighted by their four-image cross section. We do not
consider magnification bias in our statistical analysis, however we
discuss the possible consequence in Sect.4 and Sect.5. In total we
generate $\sim5\times10^6$ four-image lensing systems for the final
inspection of cusp and fold violations.

\subsection{CDM subhalos from the Aquarius and Phoenix simulations}

To populate smooth lens potentials with CDM substructures, we take two
sets of high-resolution cosmological $N$-body simulations: the
Aquarius (\citealt{volker08Aq}) and Phoenix (\citealt{Gao12Phoenix})
simulation suites. The former is composed of six Milky Way-sized halos
($M_{200}\sim10^{12}h^{-1}M_{\odot}$) and the latter consists of nine
galaxy cluster-sized halos ($M_{200}\sim10^{15}h^{-1}M_{\odot}$;
$M_{200}$ here is referred to as the virial mass, defined as the mass
within $R_{200}$, the radius within which the mean mass density of the
halo is 200 times the critical density of the Universe).

In order to estimate the lensing effects from a subhalo population
hosted by group-sized dark matter halos and to study their
dependencies on host halo properties, we rescale all these fifteen
halos from both simulations to host halos of
$M_{200}=10^{12}h^{-1}M_{\odot}, ~10^{13}h^{-1}M_{\odot},~{\rm and}~
5\times10^{13}h^{-1}M_{\odot}$. It is noteworthy to mention that we
only use $M_{200}$ of individual halos to work out their rescaling
relations, it is the subhalos (not the main halos) that we take,
rescale and then add to the constructed host lens potentials (as
described in Sect.3.1).

We define a rescaling factor $\mathfrak{R}$, which is the ratio
between $M_{200}$ of the group-sized halo and that of a simulated
halo. We rescale the masses of subhalos accordingly by a factor of
$\mathfrak{R}$, and velocities, sizes and halo-centric distances of
subhalos by a factor of $\mathfrak{R}^{\frac{1}{3}}$, so that
characteristic densities remain the same.

We also require that the minimum mass of subhalos in rescaled
group-sized halos should go down to $10^{6\sim7}h^{-1}M_{\odot}$, so
that we can analyze the lensing effects from small-scale structures
in a wide mass range; perturbations on scales below
$10^{6}h^{-1}M_{\odot}$ are not expected to alter radio flux ratios
due to finite source size effect (see detailed discussion in Xu et
al. 2012). For this reason we take both simulations at their second
resolution levels (level-2), at which the minimum resolved subhalos
have masses about seven orders of magnitude below the virial masses
of their hosts. This means that a (rescaled) halo of
$M_{200}=10^{12}h^{-1}M_{\odot}$, $10^{13}h^{-1}M_{\odot}$ and
$5\times10^{13}h^{-1}M_{\odot}$ would host a complete sample of
subhalos down to a mass of $\sim2\times10^{5}h^{-1}M_{\odot}$,
$\sim2\times10^{6}h^{-1}M_{\odot}$ and $\sim10^{7}h^{-1}M_{\odot}$,
respectively.

In Sect.4, we present results of cusp and fold violations caused by
(rescaled) subhalo populations from halos at different mass scales,
so that we can see the dependence on host halo masses. In the
following, we present the rescaled subhalo properties, including
mass function, characteristic velocities, sizes and spatial
distributions.

\subsubsection{Subhalo mass function}

From Sect.4.1 and Fig.13 of Gao et al. (2012), no significant
difference is seen between the shapes of subhalo mass functions of
cluster-sized Phoenix halos and of Milky Way-sized Aquarius halos.
The number of Phoenix subhalos is higher by 35\% than the number of
Aquarius subhalos at any fixed subhalo-to-halo mass ratio $m_{\rm
sub}/M_{200}$. This is because clusters are dynamically younger than
galaxies, therefore there are more subhalos surviving the tidal
destruction.

\subsubsection{Spatial distributions and projection effects}

\begin{figure}
\centering
\includegraphics[width=8cm]{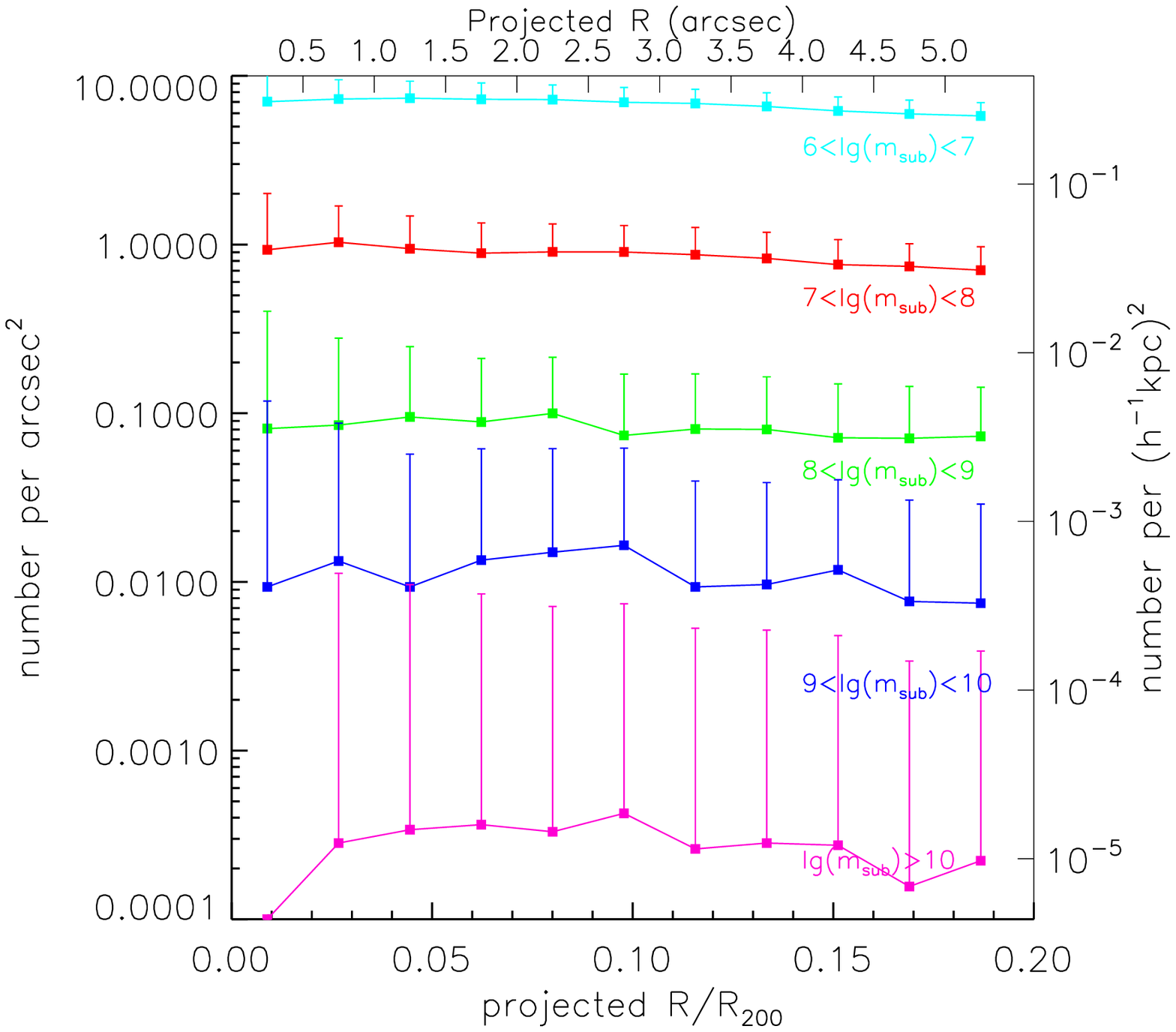}
\includegraphics[width=8cm]{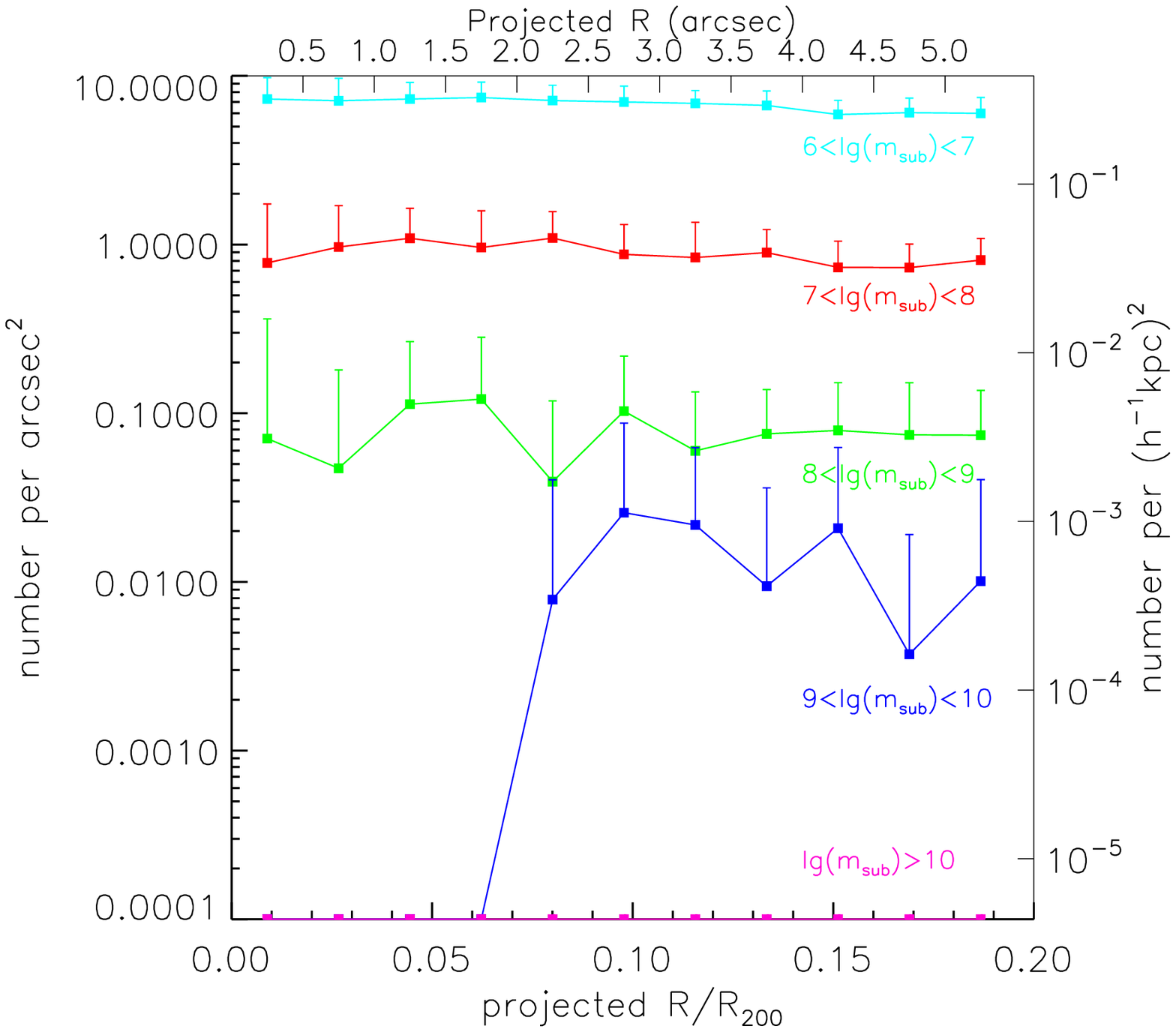}
\caption{The radial distributions of projected subhalo number
  densities from the six level-2 Aquarius halos at redshift
  $z=0.6$. All halos are rescaled to $10^{12} h^{-1}M_{\odot}$. Five
  different subhalo mass ranges have been inspected: $10^{6\sim7}
  h^{-1}M_{\odot}$ (cyan), $10^{7\sim8} h^{-1}M_{\odot}$ (red),
  $10^{8\sim9} h^{-1}M_{\odot}$ (green), $10^{9\sim10}
  h^{-1}M_{\odot}$ (blue) and $>10^{10} h^{-1}M_{\odot}$ (pink). The
  $X$-axis at the top gives the projected radius in arcsec; the one at
  the bottom gives the projected radius normalized to $R_{200}$. The
  $Y$-axis on the left gives number per arcsec$^2$; on the right gives
  number per ($h^{-1}$kpc)$^2$ (in physical scale). One-sided error bars
  indicate standard deviations. The top panel shows the result
  when 500 random projections are used per halo; the bottom panel
  when only three random projections are used per halo. }
\label{fig:AquariusProjection}
\end{figure}

\begin{figure}
\centering
\includegraphics[width=7.2cm]{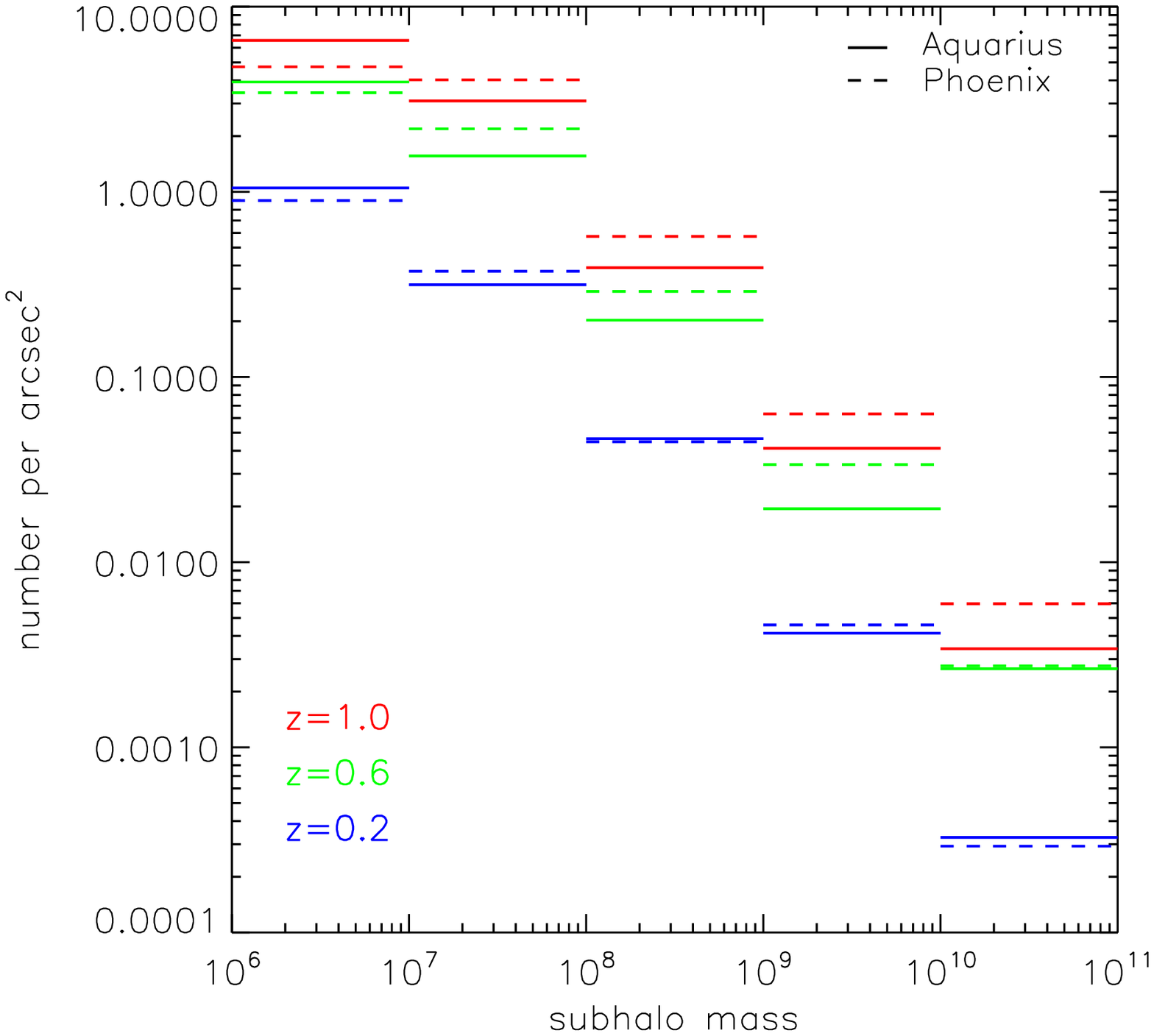}
\includegraphics[width=8cm]{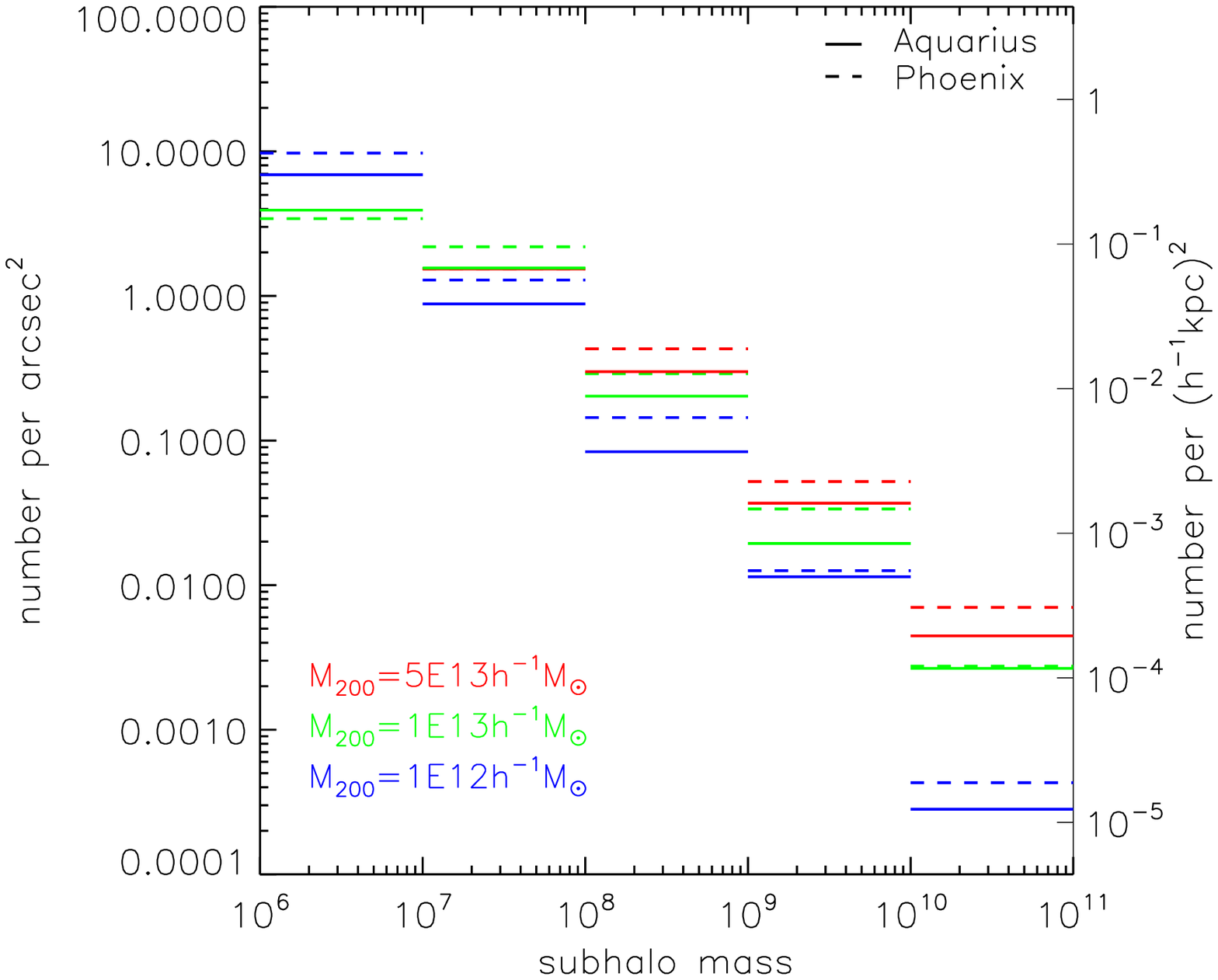}
\caption{Projected subhalo number densities averaged within the
  central $R\leqslant 5\arcsec$ region, as a function of subhalo
  masses. The top panel shows the redshift dependence: host
  halos are rescaled to $M_{200}=10^{13} h^{-1}M_{\odot}$, taken at
  $z=0.2$ (blue), $z=0.6$ (green) and $z=1.0$ (red). The bottom panel
  shows the host mass dependence: halos taken at $z=0.6$,
  rescaled to $M_{200}=10^{12} h^{-1}M_{\odot}$ (blue), $10^{13}
  h^{-1}M_{\odot}$ (green) and $5\times 10^{13} h^{-1}M_{\odot}$
  (red). 500 random projections are used per halo. The $Y$-axis on the
  left-hand side of each panel gives number per arcsec$^2$. The
  $Y$-axis on the right-hand side of the bottom panel also gives number
  per ($h^{-1}$kpc)$^2$ (in physical scale corresponding to a redshift
  at $z=0.6$). Solid lines show the number densities of the Aquarius
  subhalos; dashed lines are for the Phoenix subhalos.}
  \label{fig:SubhaloSpatial}
\end{figure}

From Sect.4.2 and Fig.15 of Gao et al. (2012), the spatial
distribution of the Phoenix subhalos is slightly more concentrated
(more abundant near the centre) than that of the Aquarius subhalos
due to the assembly bias effect, as the Phoenix simulations start
from high density regions.

For this work, the projected subhalo spatial distribution in terms
of the radial distribution of projected subhalo number density is of
particular interest, as it directly influences the lensing effects
that subhalos induce. In this subsection, we show the mean projected
spatial distributions obtained from averaging over hundreds of
projections per host halo from both simulation suites.

The projected subhalo number density does not change significantly
as a function of projected radius in the inner region of the host
halo. This is seen for both Aquarius and Phoenix subhalos.
Fig.\ref{fig:AquariusProjection} plots mean projected spatial
distributions of subhalos from six level-2 Aquaruis halos (rescaled
to $M_{200}=10^{12} h^{-1}M_{\odot}$) at redshifts $z=0.6$, where
500 and 3 random projections are taken per halo for the left and
right panels, respectively. The $Y$-axis on the left-hand side gives
number per arcsec$^2$, on the right-hand side gives number per
($h^{-1}$kpc)$^2$ (in physical scale). Error bars show the standard
deviations, and different colours indicate subhalos in different
mass bins.

From Fig.\ref{fig:AquariusProjection}, we can also see how much our
previous studies (Xu et al. 2009) could be biased from taking only
three projections of each of the six Aquarius halos.  Clearly, the
projected spatial distribution of relatively massive subhalos
($m_{\rm sub}>10^{9}h^{-1}M_{\odot}$) could be strongly affected by
small number statistics: the mean number densities drop to zero in
the inner region if only three projections per halo are used, but
will be restored if many more projections are taken per halo.
Therefore, as Metcalf \& Amara (2012) expected, our previous
conclusion could have been affected due to halo-to-halo variations.

As the projected subhalo number densities remain constant in the
inner part of a host halo, we take the mean values averaged within
the central $R\leqslant5\arcsec$ region and study their dependencies
on host halo mass and redshifts. Fig.\ref{fig:SubhaloSpatial} shows
such mean number densities as a function of subhalo mass, plotted
for host halos at different redshifts and of different $M_{200}$.
Again we can see that Phoenix halos host more subhalos than Aquarius
halos do. But more importantly, rescaling to more massive host halos
will result in a higher number density of projected subhalos; the
number per arcsec$^2$ also increases with redshift.

It can also be seen from Fig.\ref{fig:SubhaloSpatial} that, as the
subhalo mass goes down by one decade each time, there is an increase
by roughly a factor of ten in the number density of projected
subhalos, i.e., $dN/d\ln m_{\rm sub}\propto m_{\rm sub}^{-1}$. This
is expected from the subhalo mass function ($dN/dm_{\rm sub}\propto
m^{-1.9}$, Springel et al. 2008), where the logarithmic slope is
close to $-$2.0.

Another clear feature of Fig.\ref{fig:SubhaloSpatial} is the
incompleteness of rescaled subhalo populations at lower masses. In
particular, when rescaling host halos to $M_{200}=5\times 10^{13}
h^{-1}M_{\odot}$, subhalos are only complete at $m_{\rm sub} \gtrsim
10^{7}h^{-1}M_{\odot}$. In order to include subhalos between
$10^{6}h^{-1}M_{\odot}$ and $10^{7}h^{-1}M_{\odot}$ in our lensing
calculation, we adopt the following method:
for each subhalo that has a mass of $10^{7}h^{-1}M_{\odot}\leqslant
m_{\rm sub}\leqslant 10^{8}h^{-1}M_{\odot}$ and is projected in the
central strong lensing region (see Sect.3.2.3), we artificially
generate another ten subhalos each with a mass of $0.1\times m_{\rm
sub}$, projected at the same halo-centric distance but with a random
azimuthal angle.

From Fig.\ref{fig:SubhaloSpatial} we can directly read out the
projected subhalo number densities $\eta^{\ast}$ for group-sized
host halos ($M_{200}\geqslant10^{13}h^{-1}M_{\odot}$), which
satisfy:
\begin{equation}
\frac{d\eta^{\ast}}{d\ln m_{\rm sub}}\approx \big(\frac{m_{\rm
sub}}{10^{6}h^{-1}M_{\odot}}\big)^{-1}(h^{-1}\kpc)^{-2},
\end{equation}
for subhalos more massive than $10^6h^{-1}M_{\odot}$. We can then
roughly estimate the surface mass density in each mass decade to be
$\approx3\times10^{6}h^{-1}M_{\odot}(h^{-1}\kpc)^{-2}$. Consider a
typical lens system with lens and source redshifts $z_{\rm l}=0.6$
and $z_{\rm s}=2.0$, the critical surface mass density $\Sigma_{\rm
cr}\approx 3\times10^9h^{-1}M_{\odot}(h^{-1}\kpc)^{-2}$; then the
total surface mass fraction in substructures (over five mass decades
above $10^6h^{-1}M_{\odot}$) around/within the critical curve (where
local surface convergence $\kappa_{\rm cr}\equiv\Sigma/\Sigma_{\rm
cr}$ $\approx$ 0.5) is about 1\%, which is higher than 0.2-0.3\% as
estimated in Xu et al. (2009). We attribute the underestimation in
our previous study to a less abundant subhalo population in halos of
lower masses, as well as to small number statistics from the limited
number of projections used therein.

As mentioned in Sect.3.1, we take a total of 3600 different
projections of simulated subhalo distributions and add them to
generalized host lens potentials. To be precise, we take 300
projections from each of the six level-2 Aquarius halos and 200
projections from each of the nine level-2 Phoenix halos (so that the
total numbers of projections are equal for Aquarius and Phoenix
subhalo distributions). We assume the source redshift to be $z_{\rm
s}=2.0$ and take simulated subhalo populations at five different
redshifts: $z_{\rm l}=[0.2,~0.4,~0.6,~0.8,~1.0]$, which follows the
lens redshift span of the CLASS survey, i.e., from $z=0.2-1$.

We have applied (1) a flat redshift distribution for the simulated
subhalo populations, i.e., 60/40 projections per Aquarius/Phoenix
halo at each of the five redshifts; and (2) a lensing cross-section
weighted redshift distribution assuming the main lens to be a
singular isothermal sphere with velocity dispersion $\sigma_{\rm
SIS}=300$ km/s (and $z_{\rm s}=2.0$), which results in [26, 63, 79,
74, 58] projections per Aquarius halo and [17, 42, 53, 50, 38]
projections per Phoenix halos at $z=[0.2,~0.4,~0.6,~0.8,~1.0]$,
respectively. It is worth noting that in terms of flux ratio
perturbation caused by CDM substructures, there is no significant
difference between these two redshift distributions. Results
presented below are calculated for a lensing cross-section weighted
redshift distribution. These high numbers of projections ensure that
biases due to halo-to-halo variations will not affect our
conclusion.


\subsubsection{Subhalo density profiles}

The peak circular velocity $V_{\rm max}$ and the radius $r_{\rm
max}$, at which $V_{\rm max}$ is reached, are two important shape
parameters for a subhalo's density profile. As can be seen from
Fig.14 of Gao et al. (2012), the relation between $V_{\rm max}$ and
$r_{\rm max}$ is the same for the Aquarius and the Phoenix subhalos.

Springel et al. (2008) studied the (3$D$) density profile of
subhalos and found them to be well fit by Einasto profiles
(\citealt{Einasto65}) with slope parameter $\alpha=0.18$,
\begin{equation}
\rho(r)=\rho_{-2} \exp
\big(-\frac{2}{\alpha}\big[\big(\frac{r}{r_{-2}}\big)^{\alpha}-1\big]\big),
\end{equation}
where $\rho_{-2}$ and $r_{-2}$ are the density and radius at which
the local slope is $-$2. For $\alpha=0.18$, $\rho_{-2}$ and $r_{-2}$
are related to $V_{\rm max}$ and $r_{\rm max}$ by $r_{\rm
  max}=2.189~r_{-2}$ and $V^2_{\rm max}=11.19~G r^2_{-2}\rho_{-2}$,
where $G$ is the gravitational constant (see e.g., Springel et
al. 2008 for more details about fitting Einasto profiles).

From both simulation sets, instead of taking particle distributions of
subhalos for ray tracing, we take the measured $V_{\rm max}$ and
$r_{\rm max}$ for each subhalo and assume an Einasto profile with
$\alpha=0.18$. We truncate the profile at a truncation radius $r_{\rm
  trnc}$, which is set to be two times the half-mass radius $r_{\rm half}$ of the
subhalo ($r_{\rm trnc}=2r_{\rm half}$); the mass enclosed within such
a truncation radius differs from the quoted subhalo mass $m_{\rm sub}$
by less than 10\%.

To speed up the lensing calculations, only subhalos that are
projected with halo-centric distances $\mathcal{D} \leqslant R_{\rm
  ctr}+r_{\rm trnc}$, are treated as truncated Einasto profiles, where
$R_{\rm ctr}$ is the radius of a central region for strong lensing
and set to be 4.0$\arcsec$. We calculate their reduced deflection
angles and the second-order derivatives of the lens potentials, and
add them to those of smooth lenses.

For those subhalos who are projected outside the central region,
i.e., $\mathcal{D}> R_{\rm ctr}+r_{\rm trnc}$, only if they are more
massive than $10^8h^{-1}M_{\odot}$, they will be included in the
model and treated as point masses, which means they will not
contribute to the convergence $\kappa$ but only provide shear
$\gamma$ in the central region. For those below
$10^8h^{-1}M_{\odot}$ (most abundant), we (safely) exclude them from
the lensing calculations to further speed up the process, as their
overall contribution to shear at a radius of $\sim\Rein$ is
$\gamma\lesssim10^{-4}$.

We are aware that the Einasto parameter $\alpha$ could vary for
different subhaloes (\citealt{Vera-Ciro2013}), and that $r_{\rm
max}$ can not be measured as accurately as $V_{\rm max}$, especially
for lower-mass subhalos.
In reality, baryons may also play a role in shaping subhalo
concentrations. An underestimation of concentration will
underestimate cusp and fold violations. To see any potential change
in the final result due to inaccurate measurements of subhalo
profiles, we simply set $r_{\rm max}$ of each subhalo to be 0.5, 1
and 2 times its current value to carry out same lensing
calculations.

Here we verify that quantitatively there is no significant
difference in the final flux ratio probability distributions
resulting from different adoptions of $r_{\rm max}$. This simply
means that our results stably reflect the violations induced by
substructures from the simulations we choose to use, and are not a
fluke due to inaccurate estimate of subhalo profiles in simulations.
But this does not mean that density profiles (concentrations) of
subhalos will not play an important role in affecting the statistics
of flux ratio anomalies. In fact, when fundamentally different
density profiles are investigated, the violation probabilities
strongly depend on subhalo density profiles (e.g.,
\citealt{RozoZentner2006subhaloMagPert, Chen2011CuspViolation,
Xu2012LOS}), which is beyond the scope of this paper.


\section{Results}

In this section, we present the flux-ratio probability distributions
resulting from our lensing calculations for numerically generated
smooth potentials plus CDM substructures.

The numerical method we use for the lensing calculation here is the
same as in our previous studies: a grid mesh with resolution of
0.005$\arcsec$/pixel covers the lens plane, where deflection angles
and second-order derivatives of the lens potentials from the host
lens and from subhalos are calculated and tabulated onto the mesh.
The Newton-Raphson iteration method is used to find images for any
given point source; the convergence error of image positions is set
to be $0.0001\arcsec$. The adopted lens-plane resolution also
guarantees that any subhalo of $m_{\rm
sub}\geqslant10^{6}h^{-1}M_{\odot}$ will be resolved by at least one
pixel at a radius where $\kappa=0.01$, and by at least 8-10 pixels
at its half-mass radius.


\subsection{Overall flux-ratio probability distributions}


Fig.\ref{fig:CuspViolationTotal} shows probability contour maps of
conditional probabilities $P(>\Rcusp\mid \Delta\phi)$ (left),
$P(>\Rcusp\mid\theta/\Rein)$ (middle) and
$P(>|\Rfold|\mid\theta_1/\Rein)$ (right). Each panel follows the
prescription of Sect.3.1 to derive the distribution of the host
galaxy population. The top panels show the result from adopting such
smooth models;
the middle panels show results from using the smooth models plus a
subhalo population hosted by a Milky Way-sized halo of
$M_{200}=10^{12}h^{-1}M_{\odot}$; the bottom panels present results
from taking a subhalo population hosted by a group-sized halo of
$5\times10^{13}h^{-1}M_{\odot}$. More than $5\times10^6$
realizations have been calculated for each case. Observed $\Rcusp$
and $|\Rfold|$ of eight lenses in our sample are also plotted (with
error bars).

A clear mass dependence can be seen from
Fig.\ref{fig:CuspViolationTotal}: the more massive the host halos
are, the higher probabilities there are for having large $\Rcusp$
and $\Rfold$. This is expected as the number of subhalos (projected
in the inner region) increases with host halo mass (see Sect.3.2).


Another remarkable feature seen from
Fig.\ref{fig:CuspViolationTotal} is that adding substructures will
significantly change the flux-ratio probability distributions for
the image triplets/pairs that have small separations, but will not
strongly affect the distributions on larger scales. Such different
behaviors are expected, as image magnification
$\mu\approx(1-2\kappa)^{-1}$, and thus $\delta\mu/\mu\propto\mu
\delta\kappa$; $\mu\rightarrow\infty$ at around the critical curves.
A close-by image configuration (i.e., in the case of small pair
separations) suggests that the image pairs are located close to the
critical curves, where a tiny density fluctuation can introduce a
huge fluctuation in magnification.

When a perturber is located near an image which is further away from
the critical curves (i.e., in the case for larger pair separations),
it is less efficient in altering image magnification via density
fluctuation. However, it could, if massive enough, shift the image
to a new position, where the magnification is different. In this
case, standard lens models (neglecting relatively massive perturbers
if they are not luminous enough to be seen) would have difficulties
fitting image positions. This is also referred to as ``astrometric
anomaly'' (e.g., \citealt{Chen2007astrometry}).

Due to the shape of the subhalo mass function,
magnification variations due to image position shifting (caused by
relatively massive subhalos) are expected to be less frequent than
magnification perturbation resulting from local density fluctuations
(of lower-mass subhalos), which will mainly occur for image pairs
with small separations around the critical curves. This is
consistent with the fact that only a small fraction of flux anomaly
systems are also reported to have astrometric anomalies
(\citealt{Biggs2004B0128,McKean2007B2045,Sluse2012COSMOGRAIL25}).

Therefore in general, the effects of substructures on inducing
magnification perturbations and thus changing the flux-ratio
probability distributions are more prominent for image
triplets/pairs with smaller separations than for their
large-separation counterparts. This makes systems with smaller
close-pair separations to be sensitive probes of CDM substructures
via flux-ratio anomaly observations.

\subsection{Comparisons to previous studies}
In this subsection, we present comparisons to previous studies, in
particular Xu et al. (2009, 2010), and the most recent work in this
topic, i.e., Metcalf \& Amara (2012), which parameterized the CDM
subhalo populations from the Aquarius simulation and quantified
their effects on flux ratio anomalies.

In order to compare with our previous work, here we calculate the
same quantity, namely,
$P(\Rcusp\geqslant0.187|\Delta\phi\leqslant90^{\circ})$, which is
the probability for realizations that satisfy
$\Delta\phi\leqslant90^{\circ}$ to have $\Rcusp\geqslant0.187$. The
value 0.187 is the observed $\Rcusp$ for B1422, and the smallest
$\Rcusp$ among observed cusp lenses.

In our previous work, an axis ratio $q=0.8$ was adopted for the SIE
profile for smooth lens modelling. Without adding substructures, the
flux ratio $\Rcusp$ is always less than 0.2 as long as $\Delta\phi
\leqslant90^{\circ}$, which leads to
$P(\Rcusp\geqslant0.187|\Delta\phi\leqslant90^{\circ})\approx 0.0$
in the absence of substructures. Including them,
$P(\Rcusp\geqslant0.187|\Delta\phi\leqslant90^{\circ})=
10\%\sim15\%$ reflects the ``pure'' contribution from substructures
in Milky Way-sized lensing galaxies.

If we ignore higher-order multipole perturbations $a_m$ and external
shear $\gamma_{\rm ext}$, and only take SIE with $q$ drawn from 847
observed sloan galaxies (Hao et al. 2006), then the derived
$P(\Rcusp\geqslant0.187|\Delta\phi\leqslant90^{\circ})=22\%$ -- in
the absence of any substructures. Further including $a_m$ and
$\gamma_{\rm ext}$ will result in another 2\% increase for this
quantity. This means that even without adding perturbations from
subhalos to smooth lens potentials, but only using a wide range of
axis ratios/elliptiticies can also substantially raise up the value
of this statistical quantity
$P(\Rcusp\geqslant0.187|\Delta\phi\leqslant90^{\circ})$.

We note that this quantity does not increase linearly with the
amount of substructures added to the smooth potential, i.e., $P^{\rm
total}\neq P^{\rm sub}+P^{\rm smooth}$. Including substructures from
host halos of
$M_{200}=10^{12}h^{-1}M_{\odot},~10^{13}h^{-1}M_{\odot}~{\rm
and}~5\times 10^{13} h^{-1}M_{\odot}$,
$P(\Rcusp\geqslant0.187|\Delta\phi\leqslant90^{\circ})=27\%,~31\%~{\rm
and}~36\%$, respectively.

From these numbers above, it can be seen that, as Metcalf \& Amara
(2012) were concerned, the derived violation rates from our previous
studies (Xu et al. 2009, 2010) have been underestimated due to
restrictions to a small ellipticity in the lens modelling;
furthermore, using Milky Way-sized dark matter halo simulations will
indeed result in smaller violation probabilities than that for the
more massive group-sized halos (and subhalo populations therein).

Metcalf \& Amara (2012) compared flux-ratio probability
distributions in the presence of CDM substructures to an
observational sample of seven lenses and found rough consistency
between CDM simulations and observational data. Here we also present
a comparison between the probability distribution of $P(>\Rcusp\mid
\Delta\phi)$ derived in this work and that from their simulations.



Using the generic lens models therein, only one system, i.e., B2045,
out of a total of seven lenses in their observational sample was an
outlier, which was defined as lying outside the 5\% contour level of
$P(>\Rcusp\mid\Delta\phi\leqslant180^{\circ})$ in the absence of
substructures. Using our generic lens models, if we also define
outliers of the $\Rcusp$ distribution to be those outside the 5\%
contour level of $P(>\Rcusp\mid\Delta\phi\leqslant180^{\circ})$ in
the absence of any substructures, then apart from B2045, another two
lenses B0712 and B1555 in our common sample, will also be outliers
in this work.
We attribute the difference to the slightly different ranges of
smooth model parameters ($e$, $a_3$ and $a_4$, $\gamma_{\rm ext}$)
between our and their lens models.

After adding substructures, it can be seen from Fig.\,8 in Metcalf
\& Amara (2012) that for the fraction of outliers to be
$>1/7(\approx14\%)$ in the presence of CDM substructures, the
subhalo surface number density $\eta^{\ast}$ needs to be
$\gtrsim0.16~\kpc^{-2}$ at $m_{\rm sub}\geqslant10^7 M_{\odot}$, or
$\eta^{\ast}\gtrsim0.03~\kpc^{-2}$ at $m_{\rm sub}\geqslant10^8
M_{\odot}$. Such number densities were considered in their subhalo
models; however are not possible according to the simulated subhalos
that we use here. In the presence of CDM subhalos of $m_{\rm
sub}\geqslant10^7 M_{\odot}$, the fraction of outliers from our
simulations can only reach $\sim8\%$. The corresponding surface
number density $\eta^{\ast} \approx 0.05~\kpc^{-2}$ for $m_{\rm
sub}\geqslant10^7 M_{\odot}$, and $\eta^{\ast}\approx
0.005~\kpc^{-2}$ for $m_{\rm sub}\geqslant10^8 M_{\odot}$; in both
cases a factor of a few smaller than those in Metcalf \& Amara
(2012).

As we are directly counting and averaging the number of projected
subhalos using thousands of projections, we pick up the unbiased
distribution of subhalos from the simulations. We caution that the
analytical treatment in Metcalf \& Amara (2012) could be too
optimistic regarding the subhalo abundance towards the central
region of the host.

\subsection{Comparisons to our observational sample}

In this subsection, we compare the simulation results to our
observational sample. As explained in Sect.4.1, due to the behavior
of magnifications near the critical curve, we should use systems
with small close-pair separation as a safe probe to constrain
substructures in lensing galaxies. Here in particular we will only
discuss systems with (1) $\theta/\Rein$ no larger than the value
observed for B1555, and (2) $\theta_1/\Rein$ no larger than the
value observed for MG0414\footnote{Choosing the values for B1555 and
MG0414 as upper limits for the image separations is arbitrary.
However the choice is motivated by the fact that these systems have
relatively large image separations ($\theta/\Rein$ and
$\theta_1/\Rein$) but their images are still located close to the
critical curves: the close pairs have image magnifications
$|\mu|>10$. As can be seen from Fig.\ref{fig:CuspViolationTotal},
below such separations substructures will markedly change the
flux-ratio probability distributions.}.

As can be seen from the $|\Rfold|$ probability distribution in
Fig.\ref{fig:CuspViolationTotal}, among the lenses with small
$\theta_1/\Rein$, B1555 and B2045 are outliers (defined as $P<5\%$)
of the $|\Rfold|$ probability distribution in the absence of CDM
substructures. After adding substructures, their probabilities
increase to $\sim10\%$.

From the $\Rcusp$ probability distributions in
Fig.\ref{fig:CuspViolationTotal}, it can be seen that among the
lenses with small $\theta/\Rein$, B0712, B1555, B1933 and B2045 are
outliers (with $P<5\%$) of the $\Rcusp$ probability distributions in
the absences of CDM substructures.
After including substructures (expected for group-sized halos), the
probability for B0712 increases from 2\% to nearly 10\%. For B1555,
B2045 and B1933, the probabilities also markedly rise up, however,
remain $\lesssim 5\%$.

The image pairs in both B1555 and B2045 are located very close to
the critical curves and significantly magnified with $|\mu|>50$.
Such image configurations are very rare cases; their detection
suggests a significant magnification bias must be at work. This
effect has not been accounted in our statistical analysis but is
discussed in Sect.5.3. In addition, these systems harbor lenses
either with peculiar morphology or in a complex environment: B1555
seems to be highly flattened on HST images, while B2045 show a
prominent companion (\citealt{McKean2007B2045}). Our generalized
lens models that are used to calculate Fig.2 are not representative
of these extreme systems, resulting in possibly biased the
flux-ratio probability distributions.



As for B1933, although it has a seemingly small $\theta/\Rein$, the
images in question are actually located further away from the
critical curve with image magnifications $|\mu|<5$. Therefore local
density perturbations from substructures might not be the true
reason for the large discrepancy between the observed $\Rcusp$ and
the value predicted by the best-fit lens model. We discuss other
possible origins of flux anomalies in Sect.5.


There are two main limitations of using generalized smooth lens
models: first, as pointed out earlier, the sample of lenses we use
is likely biased towards large magnification systems, which are also
more likely to show flux-ratio anomalies. Since the distributions of
the lens and source populations are poorly known we could not
estimate properly the selection function of our sample and thus
calculate the expected magnification bias. However, we discuss in
Sect.5.3 the impact of applying different magnification cuts on the
probability of detecting a flux-ratio anomaly.
Overall the final statistics of violation probabilities could boost
up if taking magnification bias into account.

Second, the calculations are done for general lens models. In
reality, complicated lens environments, as well as the interplay
between dark matter and baryons in the inner parts of lensing
galaxies, can introduce complexities beyond those that our
simplified general lens models account for. When treated properly,
all these aspects are likely to induce extra density fluctuations
and thus cause magnification perturbations.


\section{Detailed investigation of each system in our lens sample}

As may be seen from the discussion in Sect.4,
generic lens models cannot always fairly sample specific lens
potentials and image configurations. For this reason, we further
investigate the effects of adding substructures to specific lens
models for our observed sample and see how substructures perturb
magnifications at the observed image positions for individual cases.

For each system, we fit the observed image positions, as well as the
positions of the lensing galaxies with standard lens modelling (see
Sect.5.1), and add to the best-fit lens potential the simulated
subhalo populations taken from redshifts close to the observed lens
redshift, and draw 200 projections per host halo so that there are
3000 projections in total from all fifteen halos from the two
simulation suites.

For each of the 3000 realizations of a given observed lens, we
generate 500 source positions within a radius of $\sim10\%$ of the
size of the tangential caustic region around the model-constrained
source position (with respect to the caustic), which results in a
total of $1.5\times10^6$ mock systems with image configurations
close to the observed ones for the final inspection.

\subsection{Macro models of the observed lenses}

We use a singular isothermal ellipsoidal plus a constant external
shear $\gamma_{\rm ext}$. A second lens, being either a satellite
galaxy or a galaxy group, will also be included in the model if its
optical/X-ray counterpart is seen in the same field (the induced shear
then may not be treated as constant). This leads to four of our
systems, namely B1422, B1608, B2045 and MG0414, being modelled
including a second lens, which is treated as a singular isothermal
sphere (SIS). Note that we use only astrometric measurements to
constrain the best lens models, i.e., positions of lensing galaxies,
and (VLBI/VLA) positions of lensed images, but we do not use image
flux ratios. Table \ref{tab:ObsSample-LensModel} lists parameters of
our standard lens models (SIE+$\gamma$+SIS) for the eight observed
lenses in our sample.

\begin{table*}
   \centering

\caption{Best SIE+$\gamma$ (N$_{\rm lens} = 1$) and SIE+SIS+$\gamma$
(N$_{\rm lens} = 2$) models for our sample:}

\label{tab:ObsSample-LensModel}

\begin{minipage} {\textwidth}

\begin{tabular}[b]{l|c|c|c|c|c|c|c|c|c|c}\hline\hline  ~~~~~~~Lens &
~$z_{\rm lens}$~ & $z_{\rm src}$ & $N_{\rm lens}$ &
~$\Rein$($\arcsec$)~ & ~~$e, \theta_{e} (\rm deg)$~~ & ~$\gamma,
\theta_{\gamma}(\rm deg)$~ &~$\Delta G($\arcsec$)$~ & $\chi^{2}$
(d.o.f.) & $ \chi^2_{\rm ima}, \chi^2_{\rm lens}$
   \\\hline
B0128+437$^1$ & 0.6 & 3.12 & 1  &  0.235 & 0.46, $-$27.72 & 0.213,
41.17 &
0.006 & 0.4 (1)& 0.0, 0.4\\
MG0414+0534$^1$ & 0.96 & 2.64 & 2  &  1.100, 0.181 & 0.22, 82.65 &
0.099,
$-$55.03 & 0.000 & 0.0 (0) & 0.0, 0.0\\
B0712+472$^{\dagger2}$& 0.41 & 1.34  & 1  & 0.699 & 0.36, $-$61.8 &
0.076, $-$13.35  & 0.028 & 2.0 (1) & 1.95, 0.06 \\
B1422+231$^{\dagger\dagger1}$ & 0.34 & 3.62 & 2  &  0.785, 4.450 &
0.21, $-$57.62
& 0.091, 77.47 & 0.000 & 0.0 (1) & 0.0, 0.0 \\
B1555+375$^{\dagger \dagger \dagger 3,4}$& 0.6 & 1.59 & 1 & 0.238 &
0.32,
81.26 & 0.143, $-$81.97 & 0.012 & 0.16 (1) & 0, 0.16\\
B1608+656$^1$ & 0.63  & 1.39 & 2  &  1.049, 0.094 & 0.84, 71.69  &
0.223,
$-$10.70 & 0.000 & 0.0 (0) & 0.0, 0.0\\
B1933+503$^{\dagger \dagger \dagger \dagger 5}$& 0.76 & 2.63 & 1 &
0.517 &
0.48, 43.51 & 0.032, 58.61 & 0.009  & 4.7 (1) & 1.2, 3.5\\
B2045+265$^1$ & 0.87 & 1.28 & 2  &  1.101, 0.032 & 0.11, 29.09  &
0.203,
$-$67.07 & 0.000 & 0.0 (0) & 0.0, 0.0\\
\hline
\end{tabular}
\\
Notes: ($\dagger$) Unrealistic lens models are obtained when the
nearby group positions of \citet{Fassnacht02B0712Group} or
\citet{Fassnacht2008Xray} are used; therefore the group is not
included in our lens modelling. ($\dagger\dagger$) This model uses
the X-ray centroid of the group by \citet{Grant2004Xray}.
($\dagger\dagger\dagger$) We assume $(z_{\rm lens}, z_{\rm src}) =
(0.6, 2.0)$ and use the galaxy position from CASTLES, $(\Delta_{\rm
gal} {\rm RA},~\Delta_{\rm gal} {\rm DEC})$ = ($-$0.185,
$-$0.150)$\pm$0.03$\arcsec$ with respect to image A.
($\dagger\dagger\dagger\dagger$) The model is based on positions of
the lensing galaxy and the lensed images 1, 3, 4, 6 only. In the
table, Col.4 gives the total number of lenses included for
modelling; Col.6 provides the best-fit amplitude and orientation of
the ellipticity; Col. 7 gives the external shear amplitude and the
position angle of the shear mass; Col.8 provides the observed
lensing galaxy position with respect to the best-fit lens position;
Col.9 gives the total $\chi^2$ of the best-fit lens model; Col.10
provides the independent contribution from the image and lens
astrometry to the total $\chi^2$. Note that flux ratios are not used
to constrain the models. References: (1)
\citealt{Sluse2012COSMOGRAIL25}; (2) \citealt{Jackson2000}; (3)
\citealt{Marlow1999B1555}; (4) Cfa-Arizona Space Telescope Lens
Survey (CASTLES, see http://cfa-www.harvard.edu/castles); (5)
\citealt{Cohn2001B1933}.
\end{minipage}
\end{table*}

\subsection{Analysis of individual lensing systems}
\begin{figure}
\centering
\includegraphics[width=8cm]{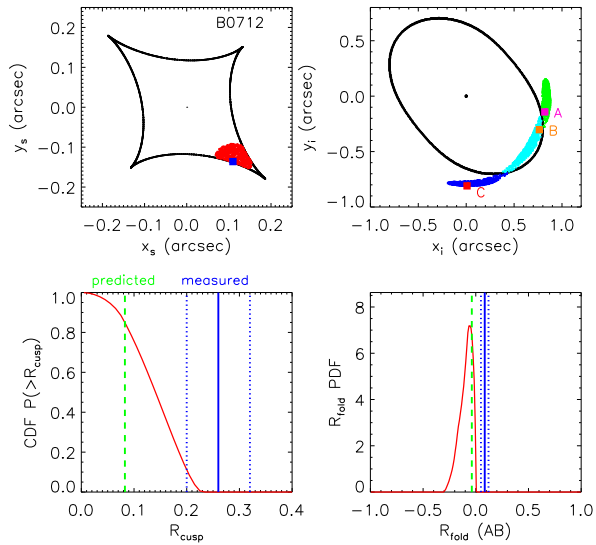}
\caption{An example of the diagnostic plots, made for B0712$+$472,
in the absence of CDM substructures. Top panels: the tangential
caustic and critical curve from the best-fit lens model. Observed
image positions of the close triplet (A, B, C) are plotted as
squares in the image plane (on the right), the predicted source
position is plotted as square in the source plane (on the left).
$1.5\times10^6$ source positions around the predicted source
position are plotted in red, the corresponding close triple images
are plotted in green, cyan and blue. Bottom panels: cumulative
probability of $\Rcusp$ larger than a given value and the
probability distribution function of $\Rfold$; measured and
predicted flux ratios are indicated with blue and green lines,
respectively. Blue dotted lines indicate error on the measurements.
} \label{fig:IndivSystemB0712Example}
\end{figure}

Fig.\ref{fig:IndivSystemB0712Example} gives an example of the
diagnostic plots, made for B0712$+$472, in the absence of CDM
substructures. Top panels show the tangential caustic and critical
curve from the best-fit lens model; observed image positions of the
close triplet are plotted in the image plane, and the corresponding
source position is plotted in the source plane. $1.5\times10^6$
source positions around the predicted source position are generated,
and the corresponding close triple images are found. The bottom
panels present the cumulative probability for $\Rcusp$ to be larger
than a given value for the close image triplet and the probability
distribution function of $\Rfold$ for the closest saddle-minimum
image pair.

It is important to note that the cumulative probability of $\Rcusp$
does not drop sharply from 1.0 to 0.0 around the model predicted
$\Rcusp$,
due to the non-local effect resulting from including a wider range
of source positions.


To select realizations that better resemble the observed systems, we
have applied stricter criteria on the image configuration parameters
$\Delta\phi$, $\theta/\Rein$ and $\theta_1/\theta_2$ of each
simulated system, where $\theta_2$ is the distance between the
second closest image pairs in the triplet configuration. We require
the relative differences between the simulated and the observed
quantities to be no larger than 10\%:
\begin{equation}
\left\{\begin{array}{c} \bigg|\frac{(\Delta\phi)_{\rm
sim}}{(\Delta\phi)_{\rm obs}}-1\bigg|\leqslant10\%, \\
\bigg|\frac{(\theta/\Rein)_{\rm sim}}{(\theta/\Rein)_{\rm obs}}-1\bigg|\leqslant10\%, \\
\bigg|\frac{(\theta_1/\theta_2)_{\rm sim}}{(\theta_1/\theta_2)_{\rm
obs}}-1\bigg|\leqslant10\%.
\end{array}\right.
\label{eq:selection}
\end{equation}

The choice of 10\% is arbitrary. We are aiming at only selecting
systems that most resemble the observed ones but also having enough
realizations to be statistically significant. The choice of 10\%
results in at least $2\times10^4$ realizations for each of the eight
observed lenses, and the probability distribution functions for
$\Rcusp$ and $\Rfold$ would not change much if using 20\% instead of
10\%.


There are two main advantages from studying individual systems via
using their specific image configurations and their own lens models
(plus CDM substructures). First, it gives us a handle to compensate
for magnification bias to a certain level. Second, it may help us to
identify other possible sources for flux anomalies apart from
density perturbations due to substructures. Below we present results
for each individual system.


\begin{figure*}
\centering
\includegraphics[width=16cm]{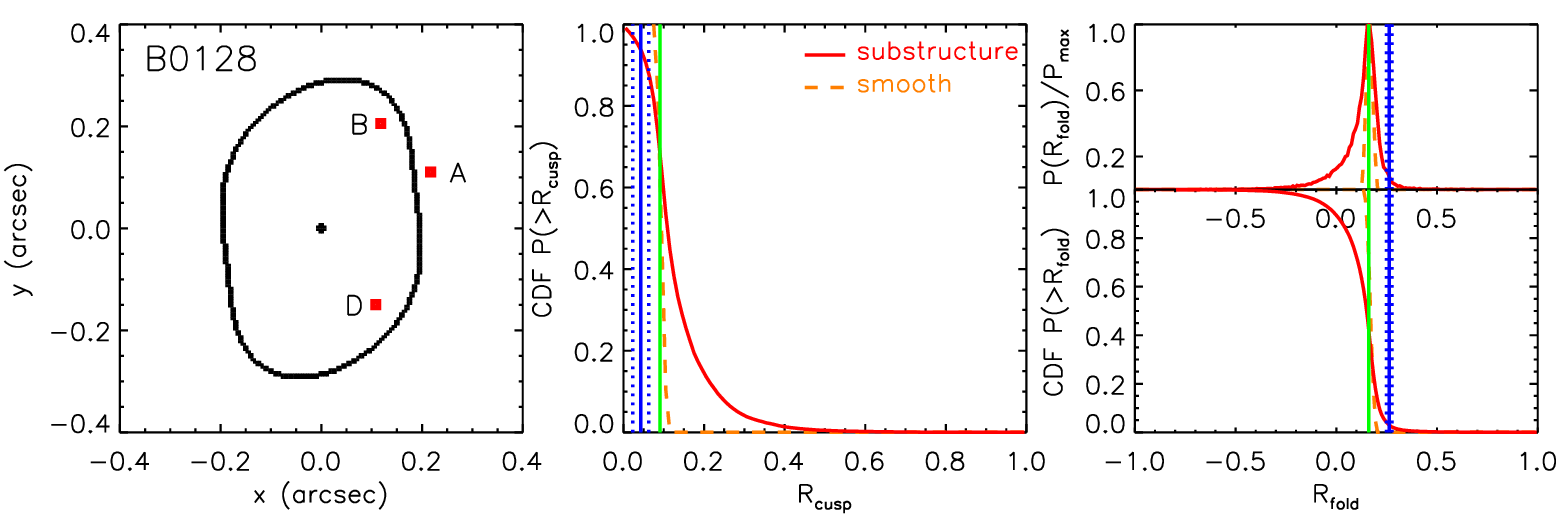}
\includegraphics[width=16cm]{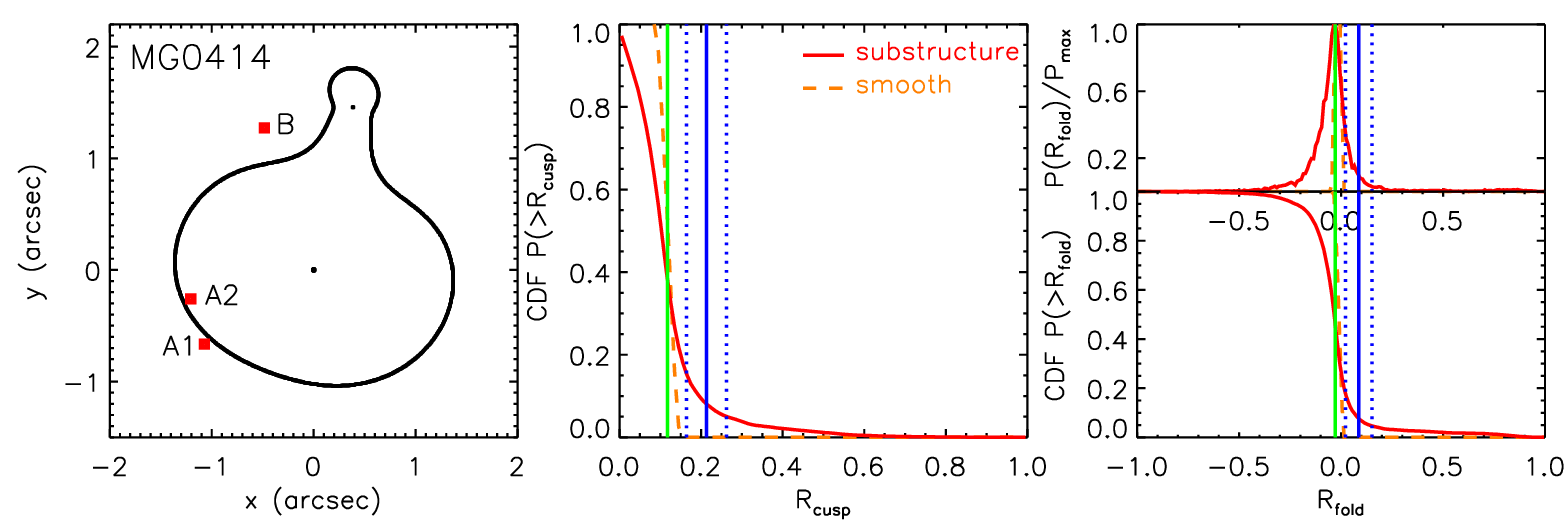}
\includegraphics[width=16cm]{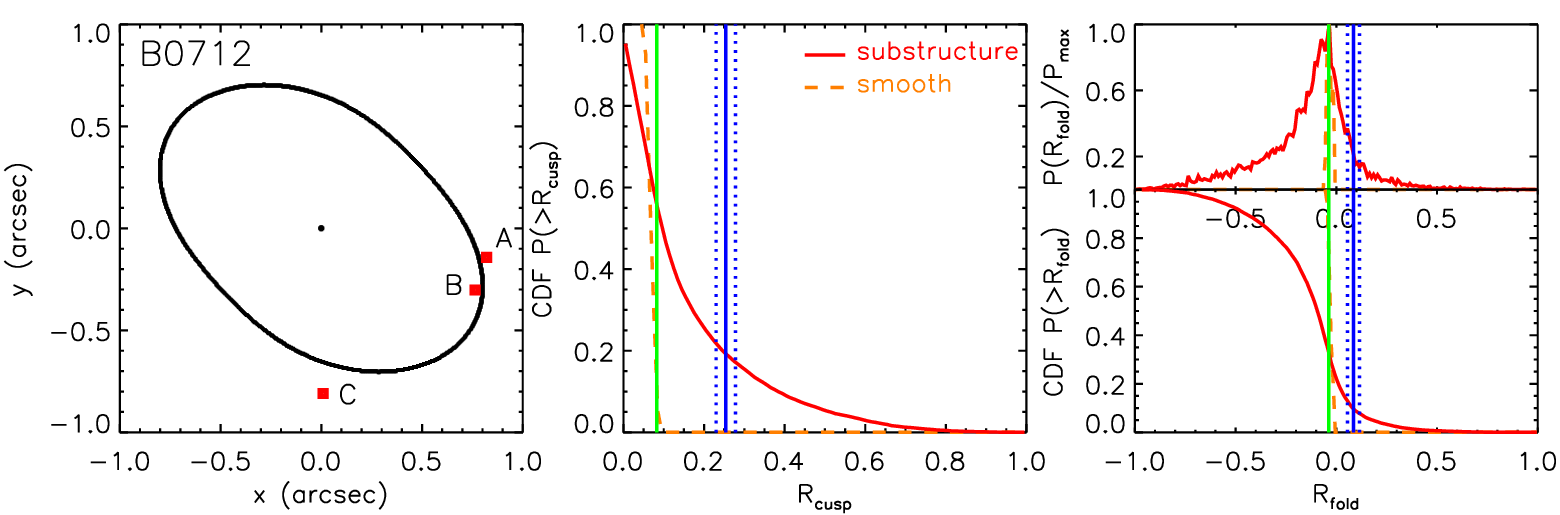}
\includegraphics[width=16cm]{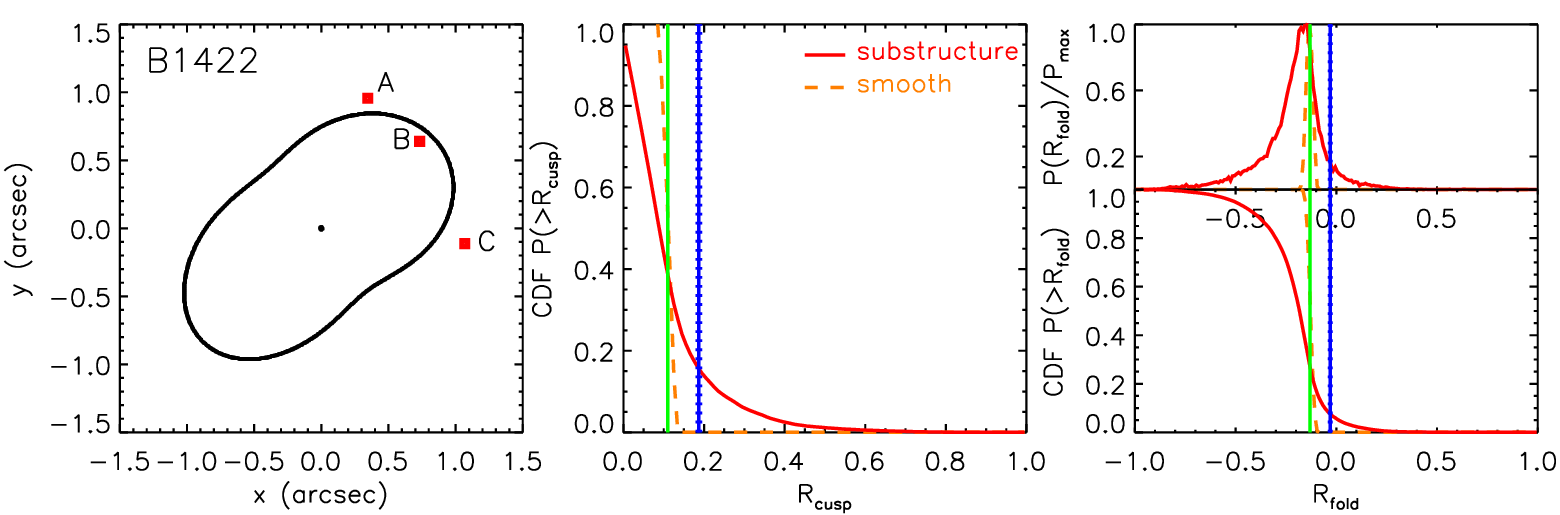}
\caption{Diagnostic plots for B0128, MG0414, B0712 and B1422, after
applying further criteria to select realizations that most resemble
the observed image configurations. Left panel: critical curves and
observed image positions; middle panel: cumulative probability
distribution functions of $\Rcusp$; right panel: cumulative and
probability distribution functions for $\Rfold$. Orange dashed lines
indicate the distribution functions calculated under macro lens
models in the absence of CDM substructures; red solid lines indicate
those after adding CDM substructures. Measured and predicted flux
ratios are indicated in blue and green lines, respectively. Blue
dotted lines indicate error on the measurements. }
\label{fig:ViolationIndivSystems1}
\end{figure*}

\begin{figure*}
\centering
\includegraphics[width=16cm]{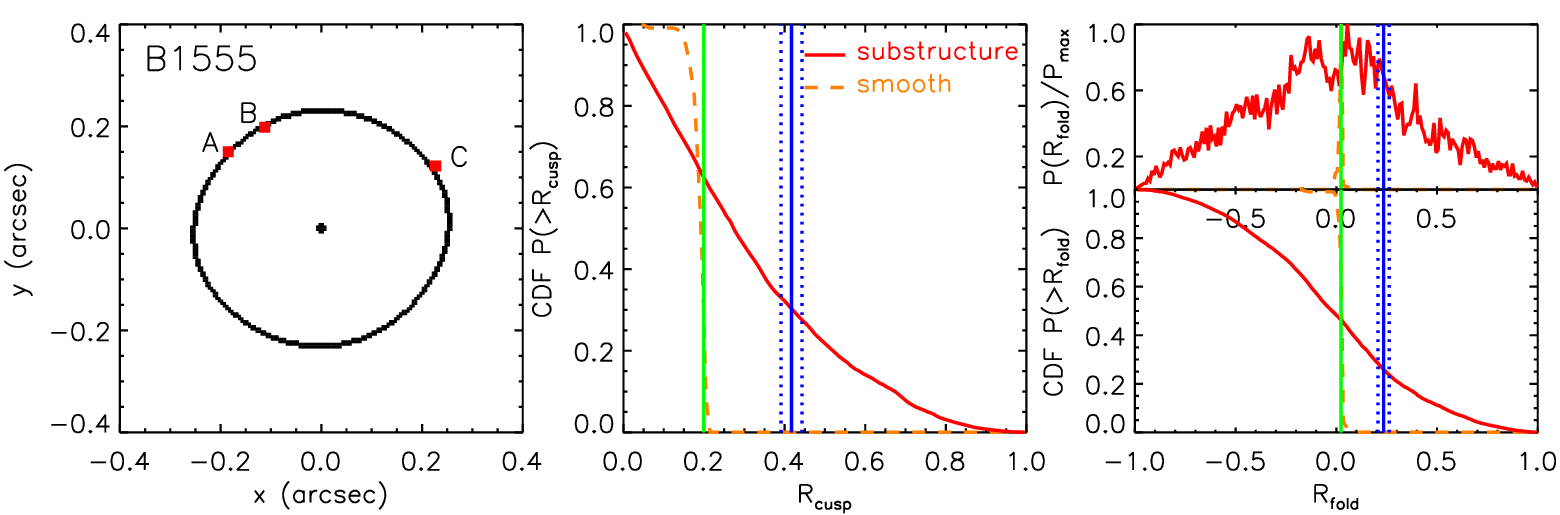}
\includegraphics[width=16cm]{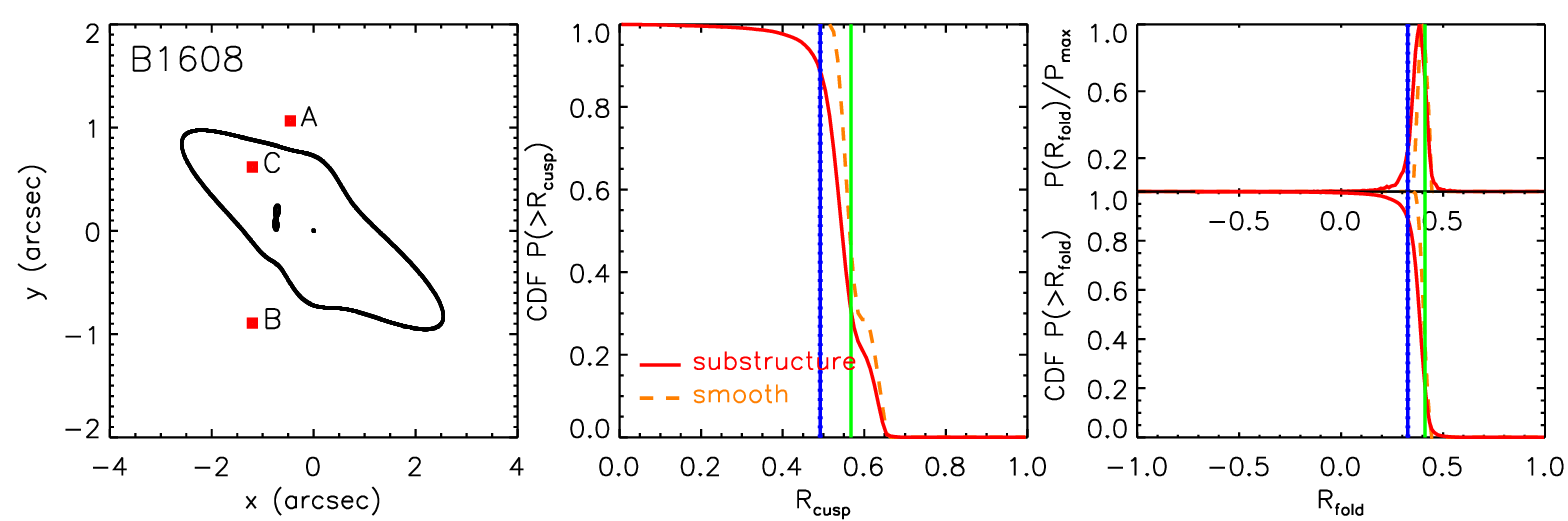}
\includegraphics[width=16cm]{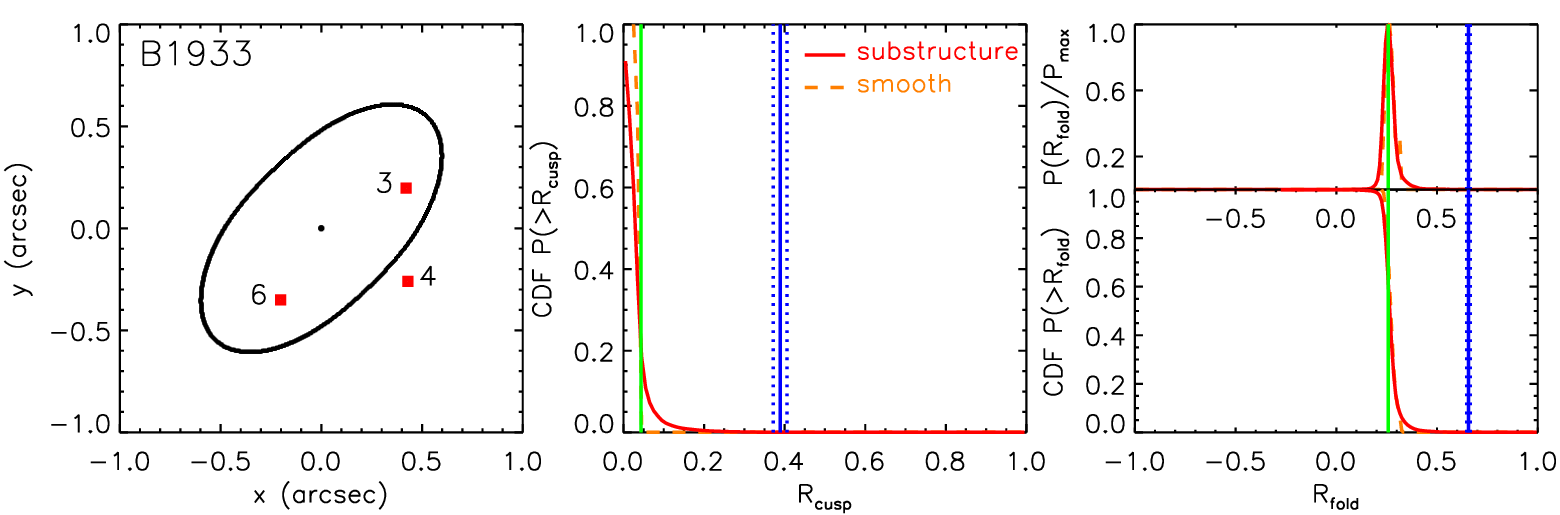}
\includegraphics[width=16cm]{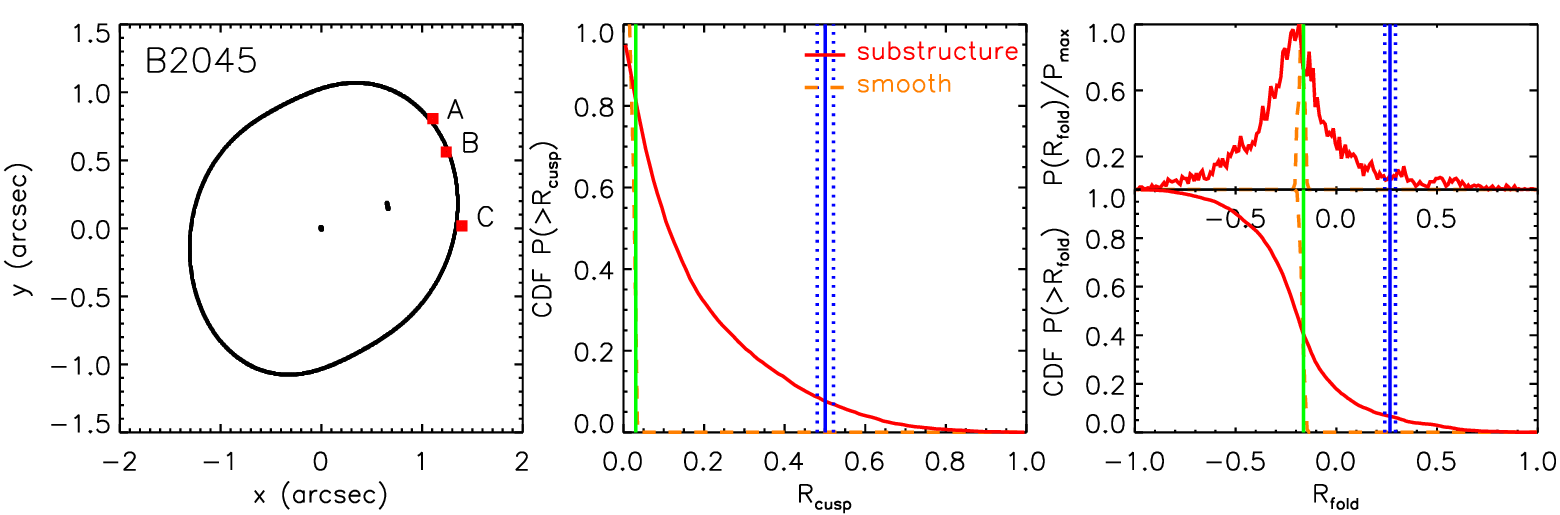}
\caption{Same as Fig.\ref{fig:ViolationIndivSystems1} but for B1555,
B1608, B1933 and B2045. } \label{fig:ViolationIndivSystems2}
\end{figure*}

\subsubsection{B0128+437}
The observed flux ratios are likely affected by complex systematic
errors, as suggested by radio-frequency dependent flux ratios and by
VLBI imaging. The VLBI data show that the source is composed of
three aligned components, one being tentatively associated with a
flat spectrum core and the other two with steep spectrum components
of the jet.
The lensed image $B$ only barely shows the ``triple'' structures,
which are visible in image $A$, $C$ and $D$. Hence it is likely that
image $B$ is affected by scatter broadening (Biggs et al. 2004).
On the other hand, lens modelling using the VLBI data also suggests
astrometric perturbation of image positions by substructures (Biggs
et al. 2004).

Despite of these uncertainties affecting the observed flux ratios,
B0128 is not an outlier of both $\Rcusp$ and $|\Rfold|$
distributions under smooth general lens models. However, our
specific lens modelling indicates that the measured and predicted
$\Rfold$ are incompatible within the measurement error. This
discrepancy is less likely to be caused by density perturbations
from substructures as the image pairs are located relatively far
from the critical curve (image magnifications $|\mu|<10$), but could
be related to the above mentioned systematic errors that affect the
observed flux ratios, or due to simplified lens modelling.


\subsubsection{MG0414+0534}

MG0414+0534 is a fold system with a pair of images close to the
critical curve (image magnifications $|\mu|>15$).  The
low-resolution radio observations of Lawrence et al. (1995) lead to
roughly the same $\Rcusp=0.36$ at epochs separated by several months
and at different frequencies from 5Ghz to 22 Ghz with the VLA,
suggesting that the time delay between the lensed images is not a
concern. However, a lower value of $\Rcusp=0.213$ was obtained from
higher-resolution VLBI observations of Ros et al. (2000), which
resolved the core+jet components of the source. The flux ratios from
VLBI for the core images also agree well with the MIR flux ratios
(\citealt{Minezaki2009MG0414}). The $\Rfold$ values from VLA, VLBI,
MIR and extinction-corrected optical data all agree with each other
within the measurement uncertainties. We use the VLBI results in
both $\Rcusp$ and $\Rfold$ for our analysis.

MG0414 is not an outlier of either $\Rcusp$ or $|\Rfold|$
distributions predicted by general lens models in the absence of CDM
substructures. However, specific lens modelling reveals
inconsistency between the measurements and model predictions. Again
the large image magnifications suggest the possibility for the flux
ratios to be (easily) affected by local density perturbations. It
can be seen that after adding CDM substructures to the macro lens
model of this system (Fig.\ref{fig:ViolationIndivSystems2}), there
is a probability of $\sim10\%$ to have $\Rcusp$ and $\Rfold$ larger
than the observed values. Interestingly,
\citet{MacLeod2012MG0414sub} have also shown that the flux ratios
between images $A1$ and $A2$ could be reproduced by adding a
substructure of $\sim 10^7 M_{\sun}$ close to image $A2$. Combining
all these pieces of evidence above, CDM substructures are highly
likely to be responsible for the observed flux anomalies in this
system.

\subsubsection{B0712+472}

B0712+472 is a cusp/fold system with a close image configuration of
$\Delta\phi=76.9^{\circ}$. The radio flux ratios seem relatively
constant over time and radio frequencies (Jackson et al. 2000,
Koopmans et al. 2003) but deviate significantly from the optical/NIR
flux ratios, which are affected by differential extinction and
microlensing (Jackson et al. 1998, 2000). This system shows
discrepancies between the observed and model predicted $\Rcusp$ and
$\Rfold$. Using general lens models, B0712 is an outlier of the
$\Rcusp$ probability distributions in the absence of substructures.

From Fig.\ref{fig:ViolationIndivSystems1} it can be seen that the
probabilities to have $\Rcusp$ and $\Rfold$ larger than the observed
values are 20\% and 10\%, respectively, again consistent with the
statistical result from Sect.4. This is strong evidence for CDM
substructures to be at least partly responsible for the observed
flux anomalies.

On the other hand, the lens environment might also impact the flux
ratios. Indeed, a galaxy group has been identified on the line of
sight towards that system (Fassnacht \& Lubin 2002, Fassnacht et al.
2008). We were unsuccessful in accounting for this group in the
smooth lens model (Table \ref{tab:ObsSample-LensModel}) due to its
uncertain X-ray centroid. A more detailed study of the lens
environment in this system is needed to estimate its effect on the
flux ratios.

\subsubsection{B1422+231}

B1422+231 is a classical cusp lens with $\Delta\phi=77^{\circ}$. The
flux ratios measured at different radio frequencies and at different
epochs and with different spatial resolution all agree with each
other, as well as with mid-infrared (MIR) data
(\citealt{Patnaik1992B1422, Patnaik1999B1422, Koopmans2003JVASCLASS,
Chiba2005}). Hence, the observed values of $\Rcusp$ and $\Rfold$ are
robustly determined. As Keeton et al. (2003) pointed out, this
system is not an outlier of the flux-ratio probability distributions
using general lens models in the absence of CDM substructures. This
can be seen again from the top panel of our
Fig.\ref{fig:CuspViolationTotal}. However, discrepancies (larger
than the measurement uncertainty) exist between measurements and
best-fit model predictions for both $\Rcusp$ of the close triple
image $A$, $B$ and $C$, and $\Rfold$ of the closest image pair $A$
and $B$ (but not for the second closest pair $B$ and $C$).

From Fig.\ref{fig:ViolationIndivSystems1} we can see that the
probability of $\Rcusp$ ($\Rfold$) larger than the observed value is
15\% (10\%) in the presence of CDM substructures, which is
consistent with the statistical result from Sect.4 and with previous
analysis of the flux ratios by \citet{Bradac2002B1422}. The
discrepancy between the measured and the predicted flux ratios are
therefore very likely to be caused by density fluctuations from CDM
substructures around image $A$, which is also consistent with
\citet{DoblerKeeton2005} using detailed lens modelling.

\subsubsection{B1555+375}
B1555+375 is a fold system, with a pair of images predicted to be
very close to the critical curve (image magnifications $|\mu|>50$).
Specific lens modellings indicate discrepancies between the observed
and model predicted $\Rcusp$ and $\Rfold$ (here and also e.g.,
Keeton et al. 2003, 2005). As discussed in Sect.4, generic cusp and
fold probability distributions would classify this system as an
outlier, even in the presence of CDM substructures.

Considering the large image magnifications, the flux ratios could be
easily affected by local density perturbations. Through specific
lens modelling in the presence of CDM substructures, the
probabilities reach as high as $\sim30\%$ for $\Rcusp$ and $\Rfold$
larger than the observed values (see
Fig.\ref{fig:ViolationIndivSystems1}). It is highly likely that
substructures are responsible for the discrepancies between
measurements and model predictions in this system.

We caution that the lens model however, might not be optimal as the
position angles of the ellipticity and of the external shear are
nearly orthogonal. The HST images of this system also suggest that
it is a very flattened lens. All these strongly indicate a possibly
missing ingredient in the lens model.
Higher resolution radio images (only MERLIN data are currently
available) as well as deep optical imaging and spectroscopy are
needed for a better characterization to this system in order to draw
firm conclusions about the macro lens model.

\subsubsection{B1608+656}
Many VLA data are available (including monitoring data) for this
system and show consistently $\Rfold \sim 0.32$. The radio
measurements of $\Rcusp$ and $\Rfold$ are larger than observed in
the optical and NIR, where the source appears to be extended and
significantly affected by differential extinction
(\citealt{Surpi2003B1608}).

For this system, both measured $\Rcusp$ and $\Rfold$ are smaller
than model predictions. As all three images in the close triplet are
located far away from the critical curve ($|\mu|<5$), we do not
expect significant magnification variation caused by local density
perturbations from CDM substructures. The main lens of B1608 is
confirmed to be a spiral galaxy (\citealt{Fassnacht1996B1608}). Such
a system may host a massive disc component, which is not
accommodated by our model but could affect flux ratios
(\citealt{Maller2000HaloDisk, Moller2003Disc}).


\subsubsection{B1933+503}
B1933+503 is also a fold system, showing discrepancies between the
observed and model predicted $\Rcusp$ and $\Rfold$. The VLBI images
presented in \citet{Suyu2012B1933} reveal that the cores in images 1
and 4 show two peaks but not for image 3. This suggests that scatter
broadening may modify the radio flux ratios.
The $\Rcusp$ and $\Rfold$ obtained from this high resolution images
also agree with lower resolution VLA and MERLIN data (Sykes et al.
1998).

On the other hand, all three images in the close triplet are located
far away from the critical curve ($|\mu|<5$), such that we do not
expect local density perturbations from CDM substructures to be
responsible for the observed flux anomalies. This can also be seen
from Fig.\ref{fig:ViolationIndivSystems2}, where CDM substructures
can do almost nothing to increase the probability for the observed
flux ratios for this system.

Overall, it seems that the uncertainties on the observed flux ratios
and the use of a too simplified macro model (as suggested by a
$\chi^2=4.7$, from our Table \ref{tab:ObsSample-LensModel}) may
account for the anomalies in this system.



\subsubsection{B2045+265}
B2045+265 is a very extreme cusp lens with
$\Delta\phi=34.9^{\circ}$. All three images are located
(symmetrically) close to the critical curve with image
magnifications $|\mu|>50$. The radio flux ratios are very robust at
different spatial resolution (VLA, VLBA) over different periods of
time, and consistent with the H-K wavelengths (Fassnacht et al.
1999, McKean et al. 2007). Koopmans (2003) identified significant
intrinsic variability at radio wavelengths, but the amplitude of
this effect is apparently small, at least on time scale of months.
This lens shows flux anomalies in both $\Rcusp$ and $\Rfold$ through
specific lens modelling. The VLBA data reveals a core+jet emission
for image $A$, but not for the saddle point image $B$, which should
be brighter than $A$ according to the models. This indicates a
possibility of the presence of a substructure around image $B$ that
demagnifies both the compact core and jet emissions.

Same as for B1555, this system would be an outlier of generic cusp
and fold probability distributions, even in the presence of CDM
substructures (see Sect.4). But the close configuration and large
image magnifications suggest that the flux ratios can be easily
affected by density perturbations from substructures. This can be
seen clearly after adding CDM substructures to the best-fit lens
potential of this system (Fig.\ref{fig:ViolationIndivSystems2}), the
probabilities now reach $\sim10\%$ for $\Rcusp$ and $\Rfold$ to be
larger than the observed values.

\subsection{Selection and magnification effects}


The flux-ratio probability distributions presented in
Fig.\ref{fig:ViolationIndivSystems1} and
\ref{fig:ViolationIndivSystems2} are calculated for realizations
with image configurations that satisfy Eq.\ref{eq:selection}. As the
true selection function is hard to quantify and thus take into
account, we caution that the statistical results may change if the
sample is selected using different criteria. Here we investigate the
flux-ratio probability distributions for realizations selected
according to image magnification instead of image configuration.

\begin{figure*}
\centering
\includegraphics[width=5.5cm]{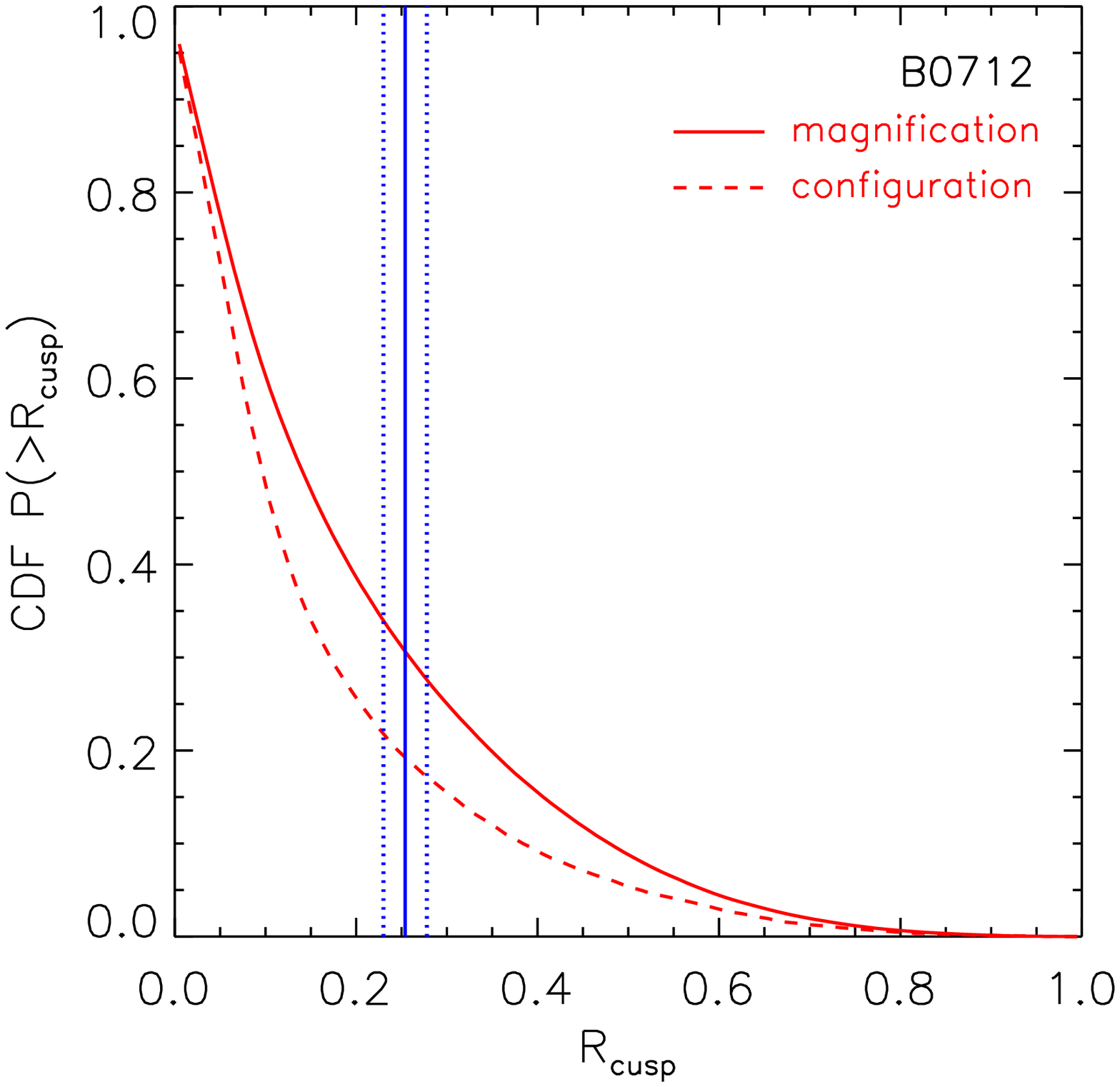}
\includegraphics[width=5.5cm]{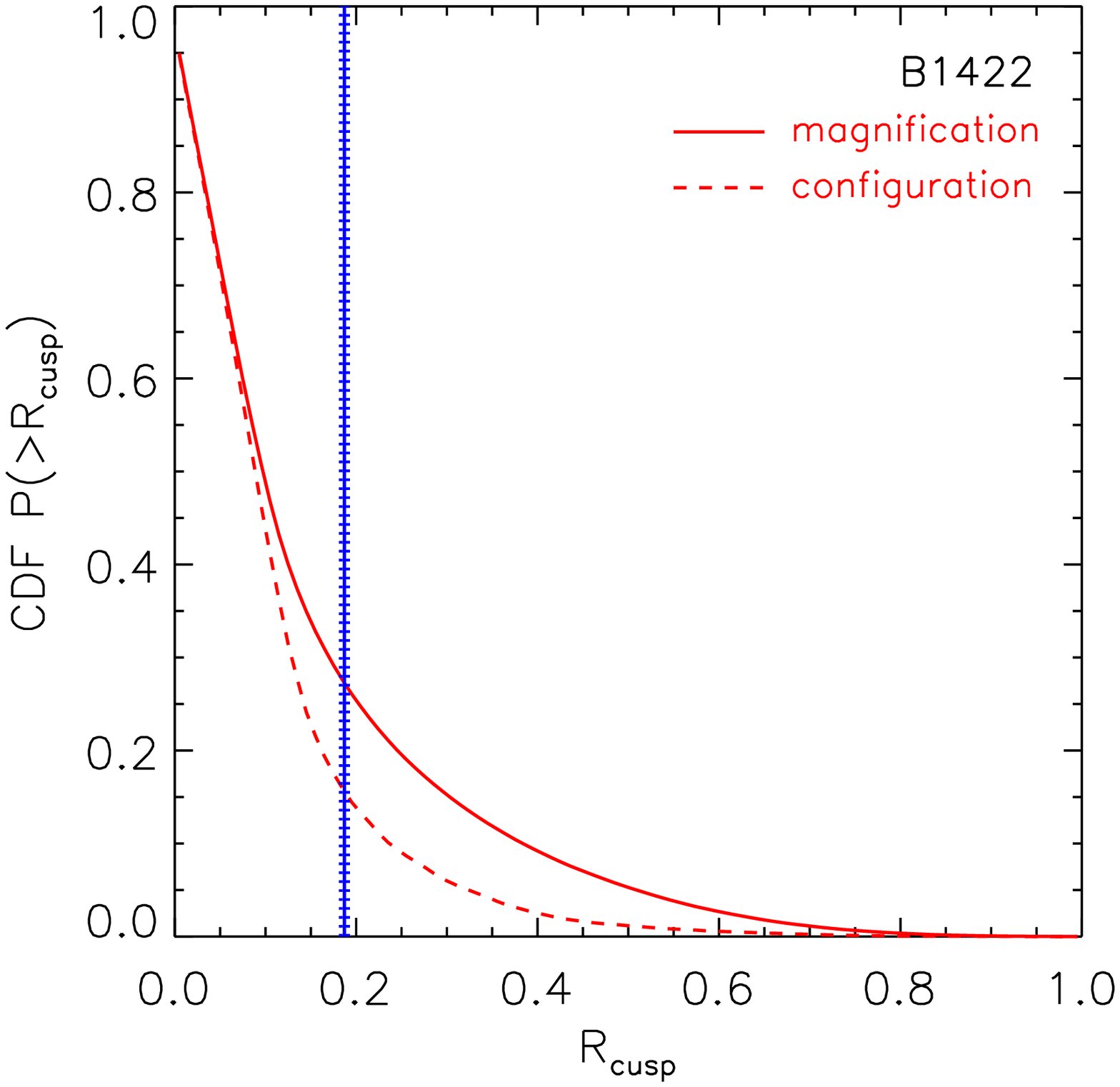}
\includegraphics[width=5.5cm]{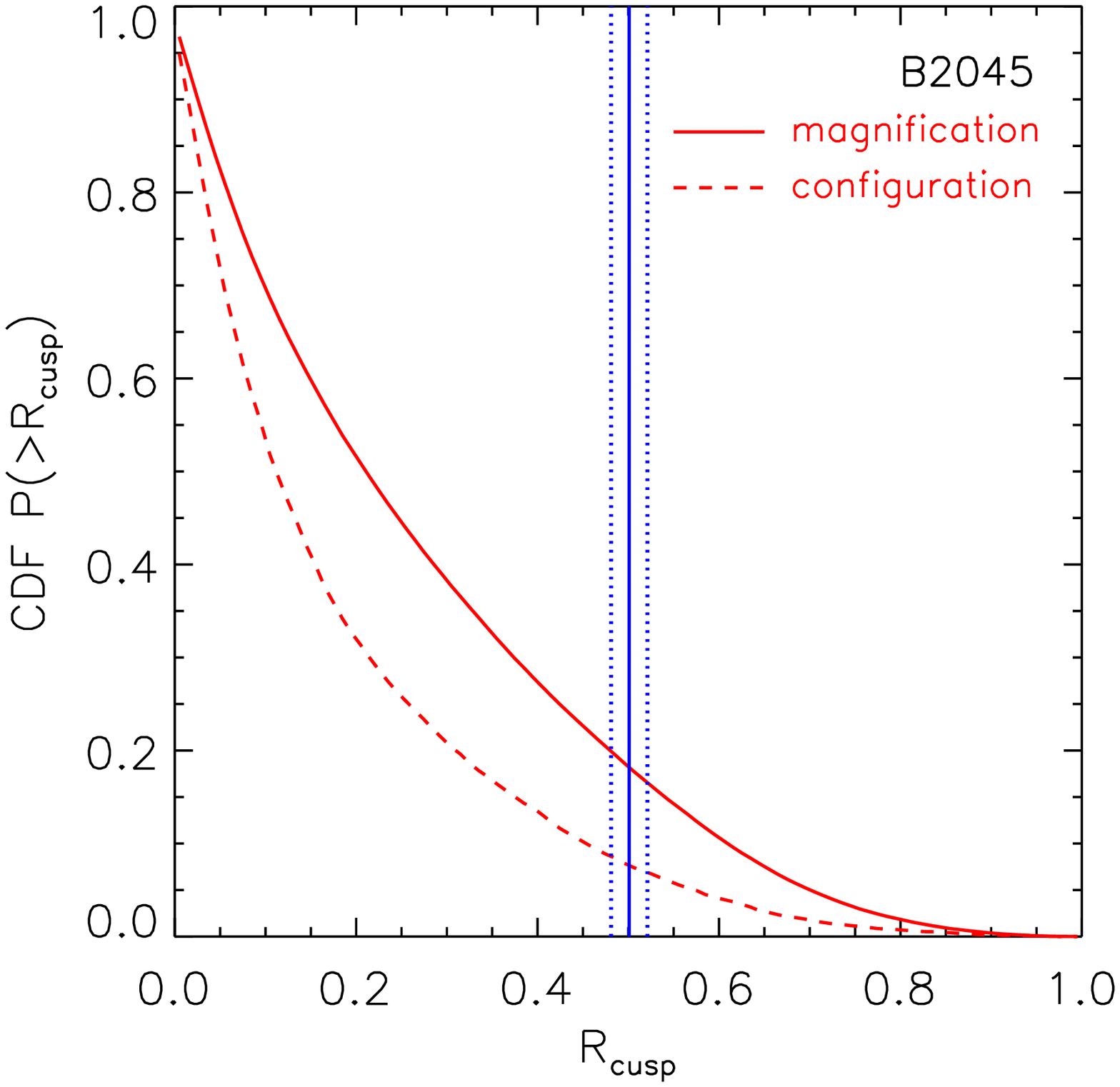}
\caption{Cumulative distribution functions of $P(>\Rcusp)$ for
system B0712, B1422 and B2045. Dashed curves are for realizations
selected according to image configurations that satisfy
Eq.\ref{eq:selection}; solid curves are for realizations whose
saddle image magnifications are no less than those in B1422, B0712
and B2045. The blue vertical lines indicate the measured $\Rcusp$
with error bars.} \label{fig:MagEffectViolation}
\end{figure*}

The final statistics are very different for B0712, B1422 and B2045;
but no significant differences are seen for the rest five systems,
as for which, the magnification selection does not lead to a
realization sample that is very different from using image
configuration selection. Fig.\ref{fig:MagEffectViolation} presents
the $\Rcusp$ probability distributions using two different criteria
to select systems that most resemble the three systems above. In
particular, the solid curves are for realizations whose saddle image
magnifications $|\mu_{\rm sad}|\geqslant20, ~15~{\rm and}~70$, which
are model predictions for B0712, B1422 and B2045, respectively.

It can be seen clearly that using realizations selected according to
magnification criteria results in higher probabilities to have
$\Rcusp$ as large as measurements. As a matter of fact, as the
applied magnification cut goes up, such probabilities will increase
accordingly. This again reflects the fact that images close to the
critical curves thus having large magnifications, are more likely to
be affected by substructures and thus show larger $\Rcusp$. Using
magnification selection criteria, the probabilities for these three
systems become $20\%\sim30\%$, which again highly suggest that CDM
substructures are responsible for the observed flux ratio anomalies
therein.

\section{DISCUSSION AND CONCLUSIONS}

Discrepancies between the observed and model-predicted flux ratios
in $\Rcusp$ and $\Rfold$ are seen in a number of radio lenses. The
interpretation of these anomalies is that substructures present in
lensing galaxies perturb the lens potentials and alter image
magnifications (and thus flux ratios). These systems have therefore
been used to constrain the subhalo abundance predicted in the CDM
model of structure formation. Our previous studies (Xu et al. 2009
and 2010) found that the number of subhalos in high-resolution
simulation of galactic CDM halos is insufficient to account for the
observed frequency of flux anomalies.

However, our previous work suffered from three shortcomings: (1) We
had access to only six high-resolution simulations which we
projected along a small number of directions; halo-to-halo
variations in the subhalo population and its projected spatial
distribution could have led to biased results; (2) Our main lens
models assumed relatively small ellipticities (axis ratios $q=0.8$)
instead of the full range and this too could have resulted in flawed
conclusions; (3) Most importantly, using the subhalo populations in
$N$-body simulations of Milky Way-sized halos could have
underestimated the flux anomaly frequency caused by the more
abundant subhalo populations of group-sized halos, which are more
likely to host the observed massive elliptical lenses.

In the first part of this work, we have attempted to overcome these
shortcomings and establish whether or not CDM substructures can
account for the flux ratio anomalies observed in the best currently
available sample of quasars (see Table \ref{tab:ObsSample-Cusp}),
nearly all of which show discrepancies between the measured $\Rcusp$
(and $\Rfold$) and those predicted by best-fit smooth models. We
assume that the general smooth lens potentials can be modelled as
isothermal ellipsoids with a wide range of axis ratios, higher-order
multipole perturbations and randomly oriented external shear
(Sect.3.1). We have analyzed two sets of state-of-the-art high
resolution CDM cosmological simulations: the Aquarius suite of
galactic halos and the Phoenix suite of cluster halos whose subhalo
populations are rescaled to those expected in group-sized halos
(Sect.3.2).

We find that each of the three shortcomings of our previous work can
indeed lead to biased results. Firstly, a limited number of halo
projections fails to fairly sample relatively massive subhalos. When
a large number of projections are used instead the true spatial
distributions are recovered in the inner parts of the host halos
(Fig.\ref{fig:AquariusProjection}).  Secondly, when a wide range of
ellipticities are considered, larger flux ratios in $\Rcusp$ and
$|\Rfold|$ are obtained even without the presence of substructures
(Sect.4.2; also Keeton et al. 2003 and 2005, Metcalf \& Amara 2012).
Thirdly, the richer subhalo populations of group-sized halos result
in significantly higher flux anomaly probabilities
(Figs.\ref{fig:CuspViolationTotal} and \ref{fig:SubhaloSpatial} and
Sect.4). We now estimate the surface mass fraction in subhalos (in
group-sized halos) within or around the tangential critical curve to
be $\sim1\%$ instead of 0.3\% as in Xu et al. (2009).

Using a generalized smooth lens population plus subhalos hosted by
group-sized halos we show in our Fig.\ref{fig:CuspViolationTotal}
and Sect.4 that substructures do not strongly change the flux-ratio
probability distributions for image triplets or pairs with large
separations; by contrast, for small separations the distributions
are significantly affected, resulting in a substantial amount of
flux-ratio variations (the anomalies). This is expected from the
magnification behavior $\mu\approx(1-2\kappa)^{-1}$, and thus
$\delta\mu/\mu\propto\mu \delta\kappa$; at around the critical
curves, $\mu\rightarrow\infty$. As a result, a tiny density
fluctuation near an image position that is close to the critical
curve will significantly perturb the local image magnification;
while a density fluctuation around an image that is further away
from the critical curve will be far less efficient in altering the
image magnification via density perturbation. For this reason, image
triplets and pairs with small separations are best probes of
substructure abundance in lensing galaxies.

The application of generic smooth lens potentials, however, has its
limitations in at least the following two aspects: (1) without
allowing for secondary lenses in the field, it cannot model systems
with complex lens environments from e.g., satellite galaxies or
nearby galaxy groups/clusters; (2) without considering magnification
effects, it cannot fairly sample systems with extreme image
configurations, for which although the corresponding source
positions only occupy a tiny fraction of the region inside the
caustic in the source plane, such rare events would be among the
brightest detections in the Universe due to huge magnification
effect.

To compensate for these limitations, in the second part of this
work, we have added CDM substructures to the model-predicted
specific lens potentials of individual systems in our sample and
studied the perturbation effects of substructures in the observed
specific image configurations; results are given in Table
\ref{tab:ObsSample-LensModel}, Fig.\ref{fig:ViolationIndivSystems1}
and \ref{fig:ViolationIndivSystems2} in Sect.5.

We have also qualitatively investigated the effect of magnifications
by applying different image magnification cuts, and compared the
results between realizations selected using magnifications cuts and
using configuration criteria (in Sect.5.3). We found that the higher
the applied magnification cut is, the larger probability there is
for the realization sample to have large $\Rcusp$, this again is
because of the behavior of the magnification perturbation:
$\delta\mu/\mu\propto\mu \delta\kappa$ around the critical curve.

Among the eight lenses in our sample, B1422, B0712, B1555, B2045 and
MG0414 have image triplets or pairs with small separations,
indicated by image magnifications $|\mu|>15$. For these systems, we
find that the probability of obtaining the observed flux ratios as a
result of perturbations by subhalos in group-sized halos are $20\%
\sim30\%$, strongly suggesting that CDM substructures can account
for a non-negligible amount of the observed flux-ratio anomalies.
This demonstrates that lensed quasars are very good probes of
substructures in distant early-type galaxies, complementary to
lensed galaxies showing surface brightness variations
(\citealt{VK2009substatistics, VKBTG2010, Vegetti2012Nature}).

%

It is important to bear in mind that, in addition to substructures
within the halo of the lensing galaxy, objects along the
line-of-sight to the lensed quasar can also perturb the lens
potential and give rise to flux-ratio anomalies (\citealt{Chen2003,
Wambsganss2005, PunchweinHilbert2009}). The contribution of these
interlopers can be as important as that of the intrinsic
substructures within the lensing galaxy \citep[e.g.][]{Metcalf2005a,
Metcalf2005b, Miranda2007, Xu2012LOS}.


Finally for systems with large close-pair image separations, e.g.,
B1933 in our sample (with image magnifications $|\mu|<5$), the
observed flux-ratio discrepancies (between the observed and
model-predicted) are unlikely due to substructure density
perturbations. Instead we attribute them either to light propagation
effects in the interstellar medium or to inaccuracies in simplified
smooth lens models. Here we make a bold prediction that applying
standard lens modelling techniques (e.g., as practiced here) to
state-of-the-art hydrodynamic simulations, in which interaction
between baryons and dark matter are taken into account, will reveal
a good fraction flux-ratio anomalies in systems with large
close-pair separations.



\section*{ACKNOWLEDGEMENTS}
This work was carried out on the DiRAC-1 and DiRAC-2 supercomputing
facilities at Durham University. DDX acknowledges the Alexander von
Humboldt foundation for the fellowship and thanks Dr. Lydia Heck for
her persistent effort maintaining the computing facilities at
Durham. DS is supported by the German \emph{Deut\-sche
For\-schungs\-ge\-mein\-schaft, DFG\/} project No. SL172/1-1. LG
acknowledges supports from the 100-talents program of the Chinese
academy of science (CAS), the National Basic Research Program of
China - Program 973 under grant No. 2009CB24901, the {\small NSFC}
grants No. 11133003, the {\small MPG} Partner Group Family and the
{\small STFC} Advanced Fellowship, and thanks the hospitality of the
Institute for Computational Cosmology (ICC) at Durham University. JW
acknowledges supports from the Newton Alumni Fellowship, the
1000-young talents program, the {\small CMST} grant No.
2013CB837900, the {\small NSFC} grant No. 11261140641, and the CAS
grant No. KJZD-EW-T01. CSF acknowledges an ERC Advanced Investigator
grant (COSMIWAY). SM thanks the CAS and National Astronomical
Observatories of CAS (NAOC) for financial support. Phoenix is a
project of the Virgo Consortium. Most simulations were carried out
on the Lenova Deepcomp7000 supercomputer of the super Computing
Center of CAS in Beijing. This work was supported in part by an STFC
Rolling Grant to the ICC.

\onecolumn
\appendix

\section{Summary of the best flux ratios for the sample of lensed systems}

We provide in Table ~\ref{tab:ObsSample} the best available flux
ratio measurements for the sample of lenses studied in the main
text. When flux ratios vary with spatial resolution due to resolved
structures in images, we provide measurements obtained at different
spatial resolution.
When available, we also report flux ratios averaged over several
epochs or corrected for time delays between images. In Table
~\ref{tab:ObsSample}, VLBA and VLBI images have typical beam sizes
of 2 ${\rm mas}^2$ while VLA and MERLIN frames have typical beam
sizes of 50 ${\rm mas}^2$.

\begin{table*}
\centering \caption{Observed lenses with measurements of $\Rcusp$
and $\Rfold$ for the close triple images:} \label{tab:ObsSample}
\begin{minipage} {\textwidth}
\begin{tabular}[b]{l|ccccccccccccccc}\hline\hline

ID & Observation & $F_1$ & $F_2$ & $F_3$ & $\Rcusp$ & ${\Rfold}$ & Image name & References \\
B0128$\dagger$ & VLA 5 Ghz 41 epochs    &  0.584$\pm$0.029 & 1.0$\pm$0.0      &  0.506$\pm$0.032 & 0.043$\pm$0.020 & 0.263$\pm$0.014 & B*-A-D*  & 1 \\
        & VLBA 5 Ghz               &  2.8$\pm$0.28 & 10.6$\pm$1.06       &  4.8$\pm$0.48    & 0.165$\pm$0.055 & 0.582$\pm$0.034 & -   & 2 \\
        & Merlin 5 Ghz             &  9.5$\pm$1   & 18.9$\pm$1           &  9.2$\pm$1       & 0.005$\pm$0.046 & 0.331$\pm$0.033 & -      & 3 \\
MG0414  & VLBI 8.5 Ghz core        & 115.6$\pm$11.56  &  97$\pm$9.7      & 34$\pm$3.4           & 0.213$\pm$0.049 & 0.087$\pm$0.065 & A1-A2*-B  & 4 \\
        & VLA 15 Ghz 4 epochs    & 157.0$\pm$5.5    &  138.75$\pm$5    & 138.75$\pm$2.25      & 0.361$\pm$0.012 & 0.062$\pm$0.024 & -      & 5 \\
        & MIR                     & 1.0$\pm$0.0      &  0.9$\pm$0.04    & 0.36$\pm$0.02        & 0.204$\pm$0.016 & 0.053$\pm$0.020 & -      & 6 \\
B0712   & VLA 5 Ghz 41 epoch    & 1.0$\pm$0.0      &  0.843$\pm$0.061 & 0.418$\pm$0.037      & 0.254$\pm$0.024 & 0.085$\pm$0.030 & A-B*-C  & 1 \\
        & VLBA 5 Ghz              & 10.7$\pm$0.15    &  8.8$\pm$0.15    & 3.6$\pm$0.15         & 0.238$\pm$0.009 & 0.097$\pm$0.010 & -    & 7 \\
B1422   & VLA 5 Ghz 41 epochs      & 1.0$\pm$0.0      &  1.062$\pm$0.009 & 0.551$\pm$0.007      & 0.187$\pm$0.004 & -0.030$\pm$0.004 & A-B*-C & 1 \\
        & VLBA 8.4 Ghz            & 152$\pm$2        &  164$\pm$2       & 81$\pm$1             & 0.174$\pm$0.006 & -0.038$\pm$0.009 & -   & 8 \\
B1555   & VLA 5 Ghz 41 epochs      & 1.0$\pm$0.0      &  0.62$\pm$0.059  & 0.507$\pm$0.073      & 0.417$\pm$0.026 & 0.235$\pm$0.028  & A-B*-C   & 1 \\
B1608$\dagger\dagger$  & VLA 8.5 Ghz   & 2.045$\pm$0.01   &  1.037$\pm$0.01  & 1.0$\pm$0.001        & 0.492$\pm$0.002 & 0.327$\pm$0.003  & A-C*-B   & 9 \\
B1933$\dagger$   & VLBA 5 Ghz    &  4.7$\pm$0.4 & 19.4$\pm$0.4         &  5.4$\pm$0.4     & 0.315$\pm$0.016 & 0.610$\pm$0.009 & 3*-4-6* & 10 \\
        & VLA 15 Ghz               &  2.5$\pm$0.4 & 15.5$\pm$0.4         &  3.2$\pm$0.4     & 0.462$\pm$0.018 & 0.722$\pm$0.009 & -      & 10 \\
B2045   & VLA 5 Ghz 41 epochs    & 1.0$\pm$0.0      &  0.578$\pm$0.059 & 0.739$\pm$0.073      & 0.501$\pm$0.020 & 0.267$\pm$0.027  & A-B*-C & 1 \\
        & VLBA 5 Ghz               & 1.0$\pm$0.01     &  0.61$\pm$0.01   & 0.93$\pm$0.01        & 0.520$\pm$0.003 & 0.242$\pm$0.007  & - & 11 \\
\hline
\end{tabular}
\\ Notes: ($\dagger$) Flux ratios are likely affected by systematic errors due to
scattering. ($\dagger\dagger$) Quoted fluxes are after correction
for the time delays. In the table, fluxes and errors (in Col.3, 4
and 5) are directly taken from literatures in their original units.
When flux errors are not available, we take 10\% of the measured
fluxes as their uncertainties. Image names (in Col.8) associated
with * indicate the images have negative parities. References (1)
Koopmans et al. 2003; (2) Biggs et al. 2004 (Table 3); (3) Phillips
et al. 2000; (4) Ros et al. 2000; (5) Lawrence et al. 1995; (6)
Minezaki et al. 2009; (7) Jackson et al. 2000; (8) Patnaik et al.
1999; (9) Fassnacht et al. 1999; (10) Sykes et al. 1998; (11) McKean
et al. 2007.
\end{minipage}

\end{table*}

\section{Generalized isothermal lens with multipole perturbation and external shear}

Consider a lens potential composed of a singular isothermal
ellipsoidal, $m^{\rm th}$-mode multipole perturbation and external
shear:
\begin{equation}
\psi(\theta,\phi)=\psi_{\rm
SIE}(\theta,\phi)+\psi_m(\theta,\phi)+\psi_{\rm ext}(\theta,\phi)
\label{eq:lenspototsum}
\end{equation}
where $\theta$ and $\phi$ are the image position
$\vec{\theta}$=($\theta_{x}$, $\theta_{y}$) in polar coordinate:
$\theta=\sqrt{\theta_{x}^2+\theta_{y}^2}$ and
$\phi=\tan^{-1}(\theta_{y}/\theta_{x})$; $\psi_{\rm SIE}$, $\psi_m$
and $\psi_{\rm ext}$ are lens potentials of an singular isothermal
ellipsoidal, $m^{\rm th}$-mode multipole perturbation and external
shear, respectively. The deflection angles and second-order
derivatives of the total lens potential are then given by:
\begin{equation}
\left\{\begin{array}{l}
\alpha_x(\theta,\phi)\equiv\frac{\partial\psi}{\partial \theta_x}=
\alpha_{\textrm{SIE},x}(\theta,\phi)+\alpha_{m,x}(\theta,\phi)
+\alpha_{\textrm{ext},x}(\theta,\phi) \\~~\\ 
\alpha_y(\theta,\phi)\equiv\frac{\partial\psi}{\partial \theta_y}=
\alpha_{\textrm{SIE},y}(\theta,\phi)+\alpha_{m,y}(\theta,\phi)
+\alpha_{\textrm{ext},y}(\theta,\phi) \\~~\\ 
\psi_{11}(\theta,\phi)\equiv\frac{\partial^2\psi}{\partial
\theta_x^2}=\psi_{\textrm{SIE},11}(\theta,\phi)+\psi_{m,11}
(\theta,\phi)+\psi_{\textrm{ext},11} (\theta,\phi) \\~~\\ 
\psi_{22}(\theta,\phi)\equiv\frac{\partial^2\psi}{\partial
\theta_y^2}=\psi_{\textrm{SIE},22}(\theta,\phi)+\psi_{m,22}
(\theta,\phi)+\psi_{\textrm{ext},22}(\theta,\phi) \\~~\\ 
\psi_{12}(\theta,\phi)\equiv\frac{\partial^2\psi}{\partial
\theta_x\theta_y}=\psi_{\textrm{SIE},12}(\theta,\phi)+\psi_{m,12}
(\theta,\phi)+\psi_{\textrm{ext},12}(\theta,\phi). 
\label{eq:firstsecondpotsum}
\end{array}\right.
\end{equation}
In our numerical approach for lensing calculations, we tabulate to a
Cartesian mesh ($\theta_x$, $\theta_y$) in the image plane values of
the reduced deflection angle and second-order derivatives of the
lens potential. Here below we give the exact analytical formulae for
the lensing quantities in Eq.\ref{eq:firstsecondpotsum}. \\

For generalized isothermal lens (plus perturbation), the lens
potential $\psi$ and convergence $\kappa$ follow the pair equations
below (Keeton et al. 2003, Appendix B2):
\begin{align}
\left\{\begin{array}{l} \psi(\theta,\phi)=\theta F(\phi) = \theta
[F_{\rm SIE}(\phi)+\sum\limits_{m=3,4}F_m(\phi)]
\\ ~~ \\ \kappa(\theta,\phi)=R(\phi)[2\theta]^{-1}=
[R_{\rm SIE}(\phi)+\sum\limits_{m=3,4}\delta
R_m(\phi)][2\theta]^{-1}. \label{eq:GIEmPair}
\end{array}\right.
\end{align}
From the Poisson equation $\nabla^2\psi=2\kappa$, $F(\phi)$ and
$R(\phi)$ are related by: $R(\phi)=F(\phi)+F^{\prime\prime}(\phi)$.
$F_{\rm SIE}(\phi)$ and $R_{\rm SIE}(\phi)$ are shape functions of a
singular isothermal ellipsoidal lens, while $F_m(\phi)$ and $\delta
R_m(\phi)$ describe the higher-order multipole perturbations. For
the generic lens model used in this work, only $m=3$ and 4 are
considered.

\subsection{Singular isothermal ellipsoidal}
Specifically, if the isothermal ellipsoidal's major and minor axes
coincide with the Cartesian axes, then the shape functions are given
by (Kassiola \& Kovner 1993; Kormann, Schneider \& Bartelmann 1994;
Keeton \& Kochanek 1998):
\begin{equation}
\left\{\begin{array}{l} R_{\rm SIE}(\phi) =
\frac{\Rein}{\sqrt{1-\epsilon\cos2\phi}}
\\ ~~ \\ F_{\rm SIE}(\phi)=\frac{\Rein}{\sqrt{2\epsilon}}\bigg[\cos\phi\tan^{-1}
\bigg(\frac{\sqrt{2\epsilon}\cos\phi}{\sqrt{1-\epsilon\cos2\phi}}\bigg)
+\sin\phi\tanh^{-1}\bigg(\frac{\sqrt{2\epsilon}\sin\phi}
{\sqrt{1-\epsilon\cos2\phi}}\bigg)\bigg]. \label{eq:SIERFpair}
\end{array}\right.
\end{equation}
where $\Rein$ is the Einstein radius of the singular isothermal
ellipsoidal, $\epsilon=(1-q^2)/(1+q^2)$ and $q\in(0,1]$ is the axis
ratio of the ellipsoidal. It can be shown that $R_{\rm SIE}(\phi)$
is the equation in polar coordinates of the ellipse at the critical
curve, where $\kappa_{\rm SIE}=\frac{1}{2}$. $R_{\rm SIE}(\phi)$
corresponds to the ellipse's equation in Cartesian coordinates:
\begin{equation}
\frac{\theta_{x}^2}{a^2}+\frac{\theta_{y}^2}{b^2} = 1,~
\textrm{where~~} a=\frac{\Rein}{\sqrt{1-\epsilon}}, ~
b=aq=\frac{\Rein}{\sqrt{1+\epsilon}}. \label{eq:isoSIECartesian}
\end{equation}
As convergence $\kappa_{\rm SIE}(\theta,\phi)=\frac{R_{\rm
SIE}(\phi)}{2\theta}$, the iso-$\kappa_{\rm SIE}$ contours follow
the ellipse $R_{\rm SIE}(\phi)$ and are scaled by $\theta^{-1}$. The
deflection angles are given by (e.g., Keeton \& Kochanek 1998):
\begin{equation}
\left\{\begin{array}{l}
\alpha_{\textrm{SIE},x}(\theta,\phi)=\frac{\Rein}{\sqrt{2\epsilon}}\tan^{-1}
(\frac{\sqrt{2\epsilon}\cos\phi}{\sqrt{1-\epsilon\cos2\phi}})
\\ ~~ \\
\alpha_{\textrm{SIE},y}(\theta,\phi)=\frac{\Rein}{\sqrt{2\epsilon}}\tanh^{-1}
(\frac{\sqrt{2\epsilon}\sin\phi}{\sqrt{1-\epsilon\cos2\phi}}).
\end{array}\right.
\end{equation}
The second-order derivatives are given by:
\begin{equation}
\left\{\begin{array}{l}
\psi_{\textrm{SIE},11}(\theta,\phi)=\frac{\Rein\sin^2\phi}{\theta
\sqrt{1-\epsilon\cos2\phi}}
\\ ~~ \\
\psi_{\textrm{SIE},22}(\theta,\phi)=\frac{\Rein\cos^2\phi}{\theta
\sqrt{1-\epsilon\cos2\phi}} \\~~\\
\psi_{\textrm{SIE},12}(\theta,\phi)=-\frac{\Rein\sin\phi\cos\phi}{\theta
\sqrt{1-\epsilon\cos2\phi}}.
\end{array}\right.
\end{equation}

\subsection{System rotation}
If the major axis of the isothermal ellipse is not aligned with the
$x$-axis, but with a position angle of $\phi_{\rm rot}$, then an
image position ($\theta,\delta$), where $\delta$ is the position
angle measured from the (positive) major axis of the isothermal
ellipse, after rotation will become ($\theta,\phi$) in the new polar
coordinates, where $\phi=\delta+\phi_{\rm rot}$. The transformation
of the lensing properties between the two systems before and after
rotation is described by:
\begin{equation}
\left\{\begin{array}{l}
\alpha_x(\theta,\phi)=\widetilde{\alpha_x}(\theta,\delta)\cos\phi_{\rm
rot}-\widetilde{\alpha_y}(\theta,\delta)\sin\phi_{\rm rot}
\\~~\\
\alpha_y(\theta,\phi)=\widetilde{\alpha_x}(\theta,\delta)\sin\phi_{\rm
rot}+\widetilde{\alpha_y}(\theta,\delta)\cos\phi_{\rm rot}
\\~~\\
\psi_{11}(\theta,\phi)=\cos^2\phi_{\rm
rot}\widetilde{\psi_{11}}(\theta,\delta)+\sin^2\phi_{\rm
rot}\widetilde{\psi_{22}}(\theta,\delta)-2\sin\phi_{\rm
rot}\cos\phi_{\rm rot}\widetilde{\psi_{12}}(\theta,\delta)
\\~~\\
\psi_{22}(\theta,\phi)=\sin^2\phi_{\rm
rot}\widetilde{\psi_{11}}(\theta,\delta)+\cos^2\phi_{\rm
rot}\widetilde{\psi_{22}}(\theta,\delta)+2\sin\phi_{\rm
rot}\cos\phi_{\rm rot}\widetilde{\psi_{12}}(\theta,\delta)
\\~~\\
\psi_{12}(\theta,\phi)=\sin\phi_{\rm rot}\cos\phi_{\rm
rot}(\widetilde{\psi_{11}}(\theta,\delta)-\widetilde{\psi_{22}}(\theta,\delta))+(\cos^2\phi_{\rm
rot}-\sin^2\phi_{\rm rot})\widetilde{\psi_{12}}(\theta,\delta)
\end{array}\right.
\label{eq:rotation}
\end{equation}
where quantities on the left side of the equations are after
rotation, those on the right and labeled with ``$\sim$'' are before
rotation. As can be seen, the rotation transformation of the
convergence $\kappa$ and shear $\gamma=\gamma_1+i\gamma_2$ in the
two systems is then:
\begin{equation}
\left\{\begin{array}{l}
\kappa(\theta,\phi)\equiv(\psi_{11}+\psi_{22})/2=\widetilde{\kappa}(\theta,\delta)
\\ ~~ \\
\gamma_1(\theta,\phi)\equiv(\psi_{11}-\psi_{22})/2=\widetilde{\gamma_1}(\theta,\delta)\cos2\phi_{\rm
rot}-\widetilde{\gamma_2}(\theta,\delta)\sin2\phi_{\rm rot}
\\ ~~ \\
\gamma_2(\theta,\phi)\equiv(\psi_{12}+\psi_{21})/2=\widetilde{\gamma_1}(\theta,\delta)\sin2\phi_{\rm
rot}+\widetilde{\gamma_2}(\theta,\delta)\cos2\phi_{\rm rot}.
\end{array}\right.
\label{eq:rotationkg}
\end{equation}

In this work, we fix the major and minor axes of the isothermal
ellipsoidal along the $x$- and $y$-axis, respectively, i.e.,
$\phi_{\rm rot}=0$. The derivations and equations given below are
for this case; the orientations of the higher-order perturbation
$\phi_m$ and external shear $\phi_{\rm ext}$ are both measured
(counter-clockwise) from to the (positive) semi-major axis of the
isothermal ellipsoidal.


\subsection{Higher-order multipole perturbations}
Now consider adding a higher-order multipole perturbation $\delta
R_m(\phi)$ to the iso-$\kappa$ ellipse $R_{\rm SIE}(\phi)$, where
$\delta R_m(\phi)$ is defined as (see Keeton et al. 2003, Appendix
B2):
\begin{equation}
\delta R_m(\phi)=a_m^{\rm pert}\cos(m(\phi-\phi_m))
\label{eq:deltaRm}
\end{equation}
where $a_m^{\rm pert}$ ($>$0) and $\phi_m$ are the amplitude and
``orientation'' of the $m^{\rm th}$-order perturbation to the perfect
ellipse $R_{\rm SIE}(\phi)$.

In a particular case for the 4$^{\rm th}$-mode perturbation, an
elliptical galaxy would be more disky if $\phi_4=0$, and more boxy if
$\phi_4=\pi/4$ (which is the same as in the conventional definition
that $\delta R_4(\phi)=a_4\cos(4\phi)$, where $a_4>0$ corresponds to a
disky galaxy and $a_4<0$ corresponds to a boxy galaxy).

From Eq.\ref{eq:GIEmPair} it can be seen that, as convergence
$\kappa(\theta,\phi)=[R_{\rm SIE}(\phi)+\Sigma\delta
R_m(\phi)][2\theta]^{-1}$, now the new iso-$\kappa$ contours follow
the perturbed ellipse $[R_{\rm SIE}(\phi)+\Sigma\delta R_m(\phi)]$ (at
$\kappa=\frac{1}{2}$) and are scaled by $\theta^{-1}$.

The corresponding shape function $F_m(\phi)$ is given by (see Keeton
et al. 2003):
\begin{equation}
F_m(\phi)=\frac{1}{1-m^2}a_m^{\rm pert}\cos(m(\phi-\phi_m)).
\label{eq:Fm}
\end{equation}
The deflection angles and second order derivatives of the potential
due to the $m^{\rm th}$ perturbation are given by:
\begin{equation}
\left\{\begin{array}{l}
\alpha_{m,x}(\theta,\phi)=F_m(\phi)\cos\phi-F^{\prime}_m(\phi)\sin\phi
\\ ~~ \\
\alpha_{m,y}(\theta,\phi)=F_m(\phi)\sin\phi+F^{\prime}_m(\phi)\cos\phi
\\ ~~ \\\psi_{m,11}(\theta,\phi)=[\theta]^{-1}\sin^2\phi[F_m(\phi)+
F^{\prime\prime}_m(\phi)]=[\theta]^{-1}\sin^2\phi\delta R_m(\phi)
\\ ~~ \\\psi_{m,22}(\theta,\phi)=[\theta]^{-1}\cos^2\phi[F_m(\phi)+
F^{\prime\prime}_m(\phi)]=[\theta]^{-1}\cos^2\phi\delta
R_m(\phi)\\~\\
\psi_{m,12}(\theta,\phi)=-[2\theta]^{-1}\sin2\phi[F_m(\phi)+
F^{\prime\prime}_m(\phi)]=-[2\theta]^{-1}\sin2\phi\delta R_m(\phi)
\end{array}\right.
\label{eq:alphaphi12Fm}
\end{equation}
where $F^{\prime}_m=\frac{\partial F_m}{\partial \phi}$ and
$F^{\prime\prime}_m=\frac{\partial^2 F_m}{\partial \phi^2}$.\\

The physical quantity of $\delta R_m$ in Eq.\ref{eq:deltaRm} is the
same as in Hao et al. (2006), where a slightly different definition is
given:
\begin{equation}
\delta R_m(\phi)=a_m\cos(m\phi)+b_m\sin(m\phi).
\label{eq:deltaRmHao}
\end{equation}
In their work, they studied the isophote shapes of 847 galaxies and
presented mean $a_m/a$ and $b_m/a$ ($a$ is the semi-major axis length
of the perfect ellipse) for $m=3,~4$, and mean ellipticity
$e(\equiv1-q)$ of the ellipse, averaged within the Petrosian
half-light radius. We use these values in our main lens
modelling. Notice that Eq.\ref{eq:deltaRmHao} can also be re-written
as: $\delta R_m(\phi)=\sqrt{a_m^2+b_m^2}\cos(m(\phi-\phi_m))$, where
$\phi_m=\frac{1}{m}\tan^{-1}(b_m/a_m)\in\frac{1}{m}[0,2\pi)$. Comparing
with Eq.\ref{eq:deltaRm}, it can be seen that:
\begin{equation}
\left\{\begin{array}{l}
a_m^{\rm pert}=\sqrt{a_m^2+b_m^2}\equiv\sqrt{(a_m/a)^2+(b_m/a)^2}\times a_{\rm SIE},
\\~~\\
\phi_m=\frac{1}{m}\tan^{-1}(b_m/a_m)
\end{array}\right.
\end{equation}
where $a_m^{\rm pert}$ is re-normalized at $\kappa=\frac{1}{2}$;
$a_{\rm SIE}=\frac{\Rein}{\sqrt{1-\epsilon}}$ as given in
Eq.\ref{eq:isoSIECartesian}. Therefore the model parameters for
higher-order perturbations ($m=3,~4$) can be fixed using
observational samples (Hao et al. 2006). \\

\subsection{Constant external shear}
The lens potential $\psi^{\rm ext}(\theta,\phi)$ caused by a
constant external shear is given by:
\begin{equation}
\psi_{\rm ext}(\theta,\phi)=-\frac{\gamma_{\rm
ext}}{2}\theta^2\cos(2(\phi-\phi_{\rm ext}))
\end{equation}
where $\gamma_{\rm ext}(>0)$ is the shear amplitude and $\phi_{\rm
  ext}\in[0,\pi)$ is the position angle of the shear mass, measured
counter-clockwise from the semi-major axis of the isothermal
ellipsoidal. External shear will not contribute to external
convergence, i.e., $\kappa_{\rm ext}=0$. The deflection angle and
second-order derivatives are given by:
\begin{equation}
\left\{\begin{array}{l}
\alpha_{\textrm{ext},x}(\theta,\phi)=-\theta\gamma_{\rm
ext}\cos(\phi-2\phi_{\rm ext})
\\~~\\ \alpha_{\textrm{ext},y}(\theta,\phi)=\theta\gamma_{\rm
ext}\sin(\phi-2\phi_{\rm ext}) \\~~\\
\psi_{\textrm{ext},11}(\theta,\phi)=-\gamma_{\rm ext}\cos(2\phi_{\rm ext}) \\~~\\
\psi_{\textrm{ext},22}(\theta,\phi)=\gamma_{\rm ext}\cos(2\phi_{\rm ext}) \\~~\\
\psi_{\textrm{ext},12}(\theta,\phi)=-\gamma_{\rm ext}\sin(2\phi_{\rm ext}).
\end{array}\right.
\end{equation}
In this work, we assume random external shear orientation in each
simulated lensing system.

\twocolumn
\bibliographystyle{mn2e}
\bibliography{ms_xudd}
\label{lastpage}

\end{document}